\documentclass[11pt]{article}
\pdfoutput=1
\usepackage[utf8]{inputenc}
\usepackage{dsfont}
\usepackage{diagbox}
\usepackage{amsfonts}
\usepackage{mathtools}
\allowdisplaybreaks[4]        
\usepackage{amssymb}
\usepackage{euscript}     
\usepackage{braket}
\usepackage{starfont}
\usepackage{color,soul}         
\usepackage{braket}
\usepackage{tensor}        
\usepackage{amsthm}
\usepackage{graphicx}
\usepackage{slashed}
\usepackage{leftidx}
\usepackage{subfigure}
\usepackage{bbm}
\definecolor{outerspace}{rgb}{0.25, 0.29, 0.3}
\definecolor{scarlet}{rgb}{1.0, 0.13, 0.0}
\usepackage[header,title,page,titletoc]{appendix}  
\definecolor{princetonorange}{rgb}{1.0, 0.56, 0.0}
\definecolor{WildStrawberry}{rgb}{1.0, 0.26, 0.64}
\definecolor{rossocorsa}{rgb}{0.83, 0.0, 0.0}
\definecolor{navyblue}{rgb}{0.0, 0.0, 0.5}
\usepackage[numbers,sort&compress]{natbib}  
\usepackage{float}
\usepackage[paper=letterpaper,margin=1in]{geometry}
\newcommand{\req}[1]{(\ref{#1})} %{Eq.\thinspace(\ref{#1})}  

\newcommand{\bea}{\begin{eqnarray}}
\newcommand{\diff}{\mathrm{d}}
\newcommand{\eea}{\end{eqnarray}}
\newcommand{\ba}{\begin{eqnarray}}
\newcommand{\ea}{\end{eqnarray}}

\newcommand{\be}{\begin{equation}}
\newcommand{\ee}{\end{equation} }
\newcommand{\beqa}{\begin{eqnarray}}
\newcommand{\eeqa}{\end{eqnarray}}
\newcommand{\beqar}{\begin{eqnarray*}}
\newcommand{\eeqar}{\end{eqnarray*}}

\renewcommand{\req}[1]{eq.~(\ref{#1})}

\newcommand{\ssc}{\scriptscriptstyle}
\newcommand{\eg}{{\it e.g.,}\ }
\newcommand{\ie}{{\it i.e.,}\ }

\newcommand{\see}{S_{\ssc \rm EE}}

\newcommand{\ctt}{C_{\ssc T}}

\DeclareMathOperator{\tr}{tr}

\def\({\left(}
\def\){\right)}
\def\[{\left[}
\def\]{\right]}

\usepackage{hyperref}
\hypersetup{
    colorlinks,
    citecolor=rossocorsa,
    filecolor=navyblue,
    linkcolor=navyblue,
    urlcolor=navyblue
}

\begin{document} 

\begin{titlepage}

\begin{center}

\phantom{ }
\vspace{3cm}

{\bf \Large{Is the EMI model a QFT?\\ An inquiry on the space of allowed entropy functions}}
\vskip 0.5cm
C\'esar A. Ag\'on${}^{\text{\Hades}}$, Pablo Bueno${}^{\text{\Zeus}}$  and Horacio Casini${}^{\text{\Kronos}}$
\vskip 0.05in
\textit{Instituto Balseiro, Centro At\'omico Bariloche}
\vskip -.4cm
\textit{ 8400-S.C. de Bariloche, R\'io Negro, Argentina}

\begin{abstract}
The mutual information $I(A,B)$ of pairs of spatially separated regions satisfies, for any $d$-dimensional CFT, a set of structural physical properties such as positivity, monotonicity, clustering, or Poincar\'e invariance, among others. If one imposes the extra requirement that $I(A,B)$ is extensive as a function of its arguments (so that the tripartite information vanishes for any set of regions, $I_3(A,B,C)\equiv 0$), a closed geometric formula involving integrals over $\partial A$ and $\partial B$ can be obtained. We explore whether this ``Extensive Mutual Information'' model (EMI), which in fact describes a free fermion in $d=2$, may similarly correspond to an actual CFT in general dimensions. Using the long-distance behavior of $I_{\rm \ssc EMI}(A,B)$ we show that, if it did, it would necessarily include a free fermion, but also that additional operators would have to be present in the model. Remarkably, we find that $I_{\rm \ssc EMI}(A,B)$ for two arbitrarily boosted spheres in general $d$ exactly matches the result for the free fermion current conformal block $G^d_{\Delta=(d-1),J=1}$. On the other hand, a detailed analysis of the subleading contribution in the long-distance regime rules out the possibility that the EMI formula represents the mutual information of any actual CFT or even any limit of CFTs. These results make manifest the incompleteness of the set of known constraints required to describe the space of allowed entropy functions in QFT. 

%lll \comment{list of Roman mythology symbols available in table 329 of \url{http://tug.ctan.org/info/symbols/comprehensive/symbols-a4.pdf}}
\end{abstract}

\end{center}

\small{\vspace{4cm}\noindent
${}^{\text{\text{\Hades}}}$cesaragon1$@$gmail.com\\
${}^{\text{\text{\Zeus}}}$pablo.bueno$@$cab.cnea.gov.ar\\
${}^{\text{\Kronos}}$casini@cab.cnea.gov.ar}

\end{titlepage}

\newpage

\tableofcontents

%%%%%%%%%%%%%%%%%%%%%%%%%%%%%%%%%%%%%%%%%%%%%%%%%%%%%%%%%%%%%%%%%
%%%%%%%%%%%%%%%%%%%%%%%%%%%%%%%%%%%%%%%%%%%%%%%%%%%%%%%%%%%%%%%%%
%%%%%%%%%%%%%%%%%%%%%%%%%%%%%%%%%%%%%%%%%%%%%%%%%%%%%%%%%%%%%%%%%
%%%%%%%%%%%%%%%%%%%%%%%%%%%%%%%%%%%%%%%%%%%%%%%%%%%%%%%%%%%%%%%%%
\section{Introduction}\label{sec:intro}

In its most straightforward formulation, quantum field theory (QFT) is defined in terms of a set of field operators acting on Hilbert space. It is well known that in the relativistic case it can be equivalently defined through correlation functions, \ie vacuum expectation values of products of fields \cite{streater2000pct}. Hence, the vacuum state and its relation to the local operator content hold the full information of the theory. The step of moving from operators to numbers turns out to be very fruitful in simplifying the mathematical description and providing resources for the elaboration of specific models, \eg the path integral formulation for Euclidean functions.    

Another formulation of QFT is based on the assignation of operator algebras to open regions of spacetime, rather than field operators at a point \cite{Haag:1992hx}. This is conceptually very natural and can be thought of as a ``basis independent'' formulation since the same algebras may be generated by many different sets of fields. One inconvenience is that the mathematical description gets more complicated, involving the theory of von Neumann algebras. In this sense, to simplify the description of models in this context, and in analogy to the field operator description, a natural idea is to assign numbers to these algebras using the vacuum state. These numbers could only represent statistical measures of vacuum fluctuations for the different regions. 

Among the different possible statistical measures, the entanglement entropy (EE) is highlighted by its simplicity, powerful properties, and connections with statistical mechanics, as well as its important role in holographic models. EE is a function of the geometry of the regions ---a ``statistical correlator''--- which can be universally defined for any QFT. The entropy has ultraviolet divergences, but this is unimportant since a simple combination of entropies for two non-intersecting regions $A, B$, called  mutual information (MI),  
\be
I(A,B)\equiv S(A)+S(B)-S(AB)\,,\label{mi}
\ee
is finite and well defined, and arguably contains the full universal information of the entropy.

 A natural question is whether the MI function contains enough information to uniquely determine the models. The answer seems to be positive. For example, it is known that for a CFT the MI for faraway regions has a series expansion in the distance from where the conformal dimensions of primary fields may be extracted \cite{Cardy.esferaslejanas,agon2016quantum}. The rest of the CFT data, the coefficient of the three point functions, may be obtained from the MI involving three or more distant regions \cite{tobe}. 

A different but related question is: what are the rules for a function $I(A,B)$ to be admissible as the MI of a QFT? Surprisingly little is known about this important question.  
A simple list is the following. We recall that $I(A,B)$ is a function of causally closed,\footnote{A region is causally closed (or a ``diamond-shaped set'') if it is the set of all points spatial to another region.} spatially separated regions $A,B$ in Minkowski space. Defining the tripartite information function 
\be
I_3(A,B,C)=I_3(A,C,B)\equiv I(A,B)+I(A,C)-I(A,BC)\,,\label{ii3}
\ee
 calling $\Lambda$ to a Lorentz transformation, and taking $X$, $Y$, as regions with spatial boundary in a common null plane, we have 
\bea
&& I(A,B) \ge 0\,,  \hspace{10.2cm}\textrm{(positivity)} \label{pos}\\
&& B\subseteq C \implies I(A,B)\le I(A,C)\,,  \hspace{6.3cm}\textrm{(monotonicity)} \label{mon}\\
&& I(A,B)=I(B,A)\,, \hspace{.5cm}I_3(A,B,C)=I_3(B,C,A)\,,\hspace{4cm} \textrm{(symmetry)} \label{sim}\\
&& I(A,B)=I(\Lambda A+x, \Lambda B +x )\,, \hspace{5.5cm} \textrm{(Poincar\'e invariance)}\label{poin}\\
&& \lim_{|x|\rightarrow \infty} I(A,B+x)=0\,, \hspace{8.5cm} \textrm{(clustering)} \\ \label{marko}
&& I(A,X)+I(A,Y)\le I(A,X\cap Y)+I(A,X\cup Y)\,.\hspace{2.9cm} \textrm{(Markov property)}  
\eea

Positivity and monotonicity are properties of MI as a measure of correlations, monotonicity encoding strong subadditivity of the entropy \cite{ohya2004quantum}. The symmetry properties (\ref{sim}) tell that MI is in fact a combination of (undefined) single region functions as in (\ref{mi}). Poincar\'e invariance and clustering follow the same properties of ordinary correlators. The Markov property  (\ref{marko}) expresses strong superadditivity of mutual information when two regions $X,Y$ have boundary on a null plane. This follows from the Markov property of entropy in a null plane and strong subadditivity \cite{casini2021mutual}. Given Lorentz invariance and certain additional (natural) continuity assumptions, the entropy is Markovian on the null plane \cite{casini2017modular,casini2018all}, so (\ref{marko}) is probably an unavoidable consequence of (\ref{mon}), (\ref{sim}) and (\ref{poin}). This Markov property has been shown to imply unitarity bounds in \cite{casini2021mutual}.   

Any QFT gives a solution to this set of axioms. But we also have solutions in dimension $d$ by dimensional reduction from theories in higher dimensions. To eliminate these and fix $d$-dimensional QFT we can further impose 
that for two disjoint regions $A, B$ sharing a planar boundary with normal $\eta$
\be \label{area}
\hspace{1cm} I(A,B+\epsilon\,\eta )\sim \epsilon^{-(d-2)}\,, \hspace{.4cm} \epsilon\rightarrow 0\,. \hspace{6.7cm} \textrm{(area law)}
\ee
In studying what other properties are needed we can search for solutions to this set of properties and see if the result is physically sensible. If it is not, the hope is that it would give us a hint of what is missing. One striking difference with the case of operator correlations is the two inequalities (\ref{pos}), (\ref{mon}), instead of an infinite tower of inequalities. This is more so considering that R\'enyi entropies of integer index $n$ (the entropy is the limit $n\rightarrow 1$) do indeed obey an infinite tower of inequalities \cite{casini2010entropy}.   

It is not easy to obtain solutions in a direct way. A notable example with a purely geometric solution is given by the holographic EE \cite{ryu2006holographic,Ryu:2006ef,Hubeny:2007xt,Lewkowycz:2013nqa,Faulkner:2013ana}. Another one is the extensive mutual information model (EMI) \cite{Casini:2008wt}. This last model simplifies the symmetry requirement by imposing 
\be
I_3(A,B,C)\equiv 0\,.\label{i3}
\ee
 According to the definition of $I_3$, eq. (\ref{ii3}), this implies an extensivity or additivity of the mutual information as a function of its arguments, and hence the name of the model.   
If we further impose conformal invariance there is a unique solution (except for a global multiplicative constant). This is given by 
\be
I_{\rm \ssc EMI}(A,B)=2 \kappa_{(d)}\, \int_{\Sigma_A} \diff \sigma_A\, \int_{\Sigma_B} \diff \sigma_B \, \eta_A^\mu(x_A) \,\eta_B^\nu(x_B) \,(\partial_\mu \partial_\nu-g_{\mu\nu} \partial^2) \,|x_A-x_B|^{-2(d-2)}\,, \label{aa}
\ee
where $\Sigma_A$, $\Sigma_B$, are any Cauchy surfaces for the causal regions $A$, $B$; $\eta_A$, $\eta_B$, are the unit normals to these surfaces, and $\kappa_{(d)}$ is a positive constant. The integrand is a conserved current in both indices, which guarantees the result to be independent of  the Cauchy surface. In fact, this expression is equivalent to one where the integration is only on the spatial boundaries $\partial A$, $\partial B$, of the regions:
%\be
%I_{\rm \ssc EMI}(A,B)\propto \,\int_{\partial A} \diff \sigma_A^{\alpha_1\cdots \alpha_{d-2}}\, \int_{\partial B} \diff \sigma^B_{\alpha_1\cdots \alpha_{d-2}}\, \frac{1}{|%x_A-x_B|^{2(d-2)}}\,,\label{sis}
%\ee
%where we integrate over the normalized volume $(d-2)$-forms on the two surfaces with the outward pointing orientation. Yet another equivalent form follows from using the %Hodge duals of the volume forms, 
\be
I_{\rm \ssc EMI}(A,B)=2\kappa_{(d)}\,\int_{\partial A} \diff \sigma_A  \, \int_{\partial B} \diff \sigma_B \, \frac{(n_A\cdot n_B )(\bar n_A \cdot \bar n_B)-(n_A\cdot \bar{n}_B )(\bar n_A \cdot n_B)}{|x_A-x_B|^{2(d-2)}}\,,\label{sis2}
\ee
where % with $\diff \sigma_x \,\omega_{\mu\nu}(x) \,\diff \sigma_y \,\omega^{\mu\nu}(y)$, where  $\omega_{\mu\nu}=n_\mu \bar n_\nu-n_\nu \bar n_\mu$, for two unit vectors 
$\bar n_A$ and $ n_A$ are unit vectors orthogonal to the surface and to each other, $n_A\cdot \bar n_A=0$.\footnote{The factor $2$ in this formula is conventional, and devised to get the same formula with $\partial A=\partial B$ for the entropy $S(A)$, with $\kappa_{(d)}$ as a coefficient.} For fixed time slices, one can choose $n_A= n_B=\hat t$ and the integrand reduces to $-(\bar n_A\cdot \bar n_B)/|x_A-x_B|^{2(d-2)}$. %These are the Hodge duals of the volume forms.

%{\color{blue} Aca chequear constantes (convencion de uso posterior), signos, convencion de la metrica. Debo prueba de igualdad de las dos formulas.}

The EMI model gives physically reasonable results, and has been mainly used to explore the expected behavior of the EE  in different geometric situations \cite{casini2015mutual,Bueno:2015rda,Witczak-Krempa:2016jhc,Bueno:2019mex,Estienne:2021lxh}. In general, the results are quite similar to the free fermion ones, and the EMI model indeed coincides with a free massless fermion in $d=2$ \cite{casini2005entanglement}. Another way to arrive at the EMI formula is to assume the R\'enyi twist operators are exponentials of free fields \cite{swingle2010mutual}. This automatically gives the bilinear form on the boundaries as in (\ref{sis2}) from the expectation values of the R\'enyi operators.  

In this paper, we investigate if the EMI model gives the mutual information of some CFT in $d\ge 3$. In other words, we test the consistency of imposing eq. (\ref{i3}) in QFT.  To understand the possible theory behind (\ref{sis2}), we use known relations between the behavior of mutual information at long distances and the operator content of CFTs. We find that no possible CFT or limit of CFTs is consistent with the extensivity condition. This result makes manifest that the list of known constraints or axioms satisfied by the MI of spatially separated regions in QFT is incomplete.  

A more detailed list of our findings can be found next.

%e urgent need for additional constraints/axioms satisfied by all QFTs beyond the known ones.

\subsection{Summary of results}
%Let us summarize here the main findings of the paper. 

In section \ref{long-dist}, we show that in case the EMI formula describes the MI of an actual CFT, this necessarily contains a free fermion. This is done by considering the leading contribution to the long-distance behavior of the EMI in the case of two boosted spheres, which reveals the same tensorial structure as in the case of a free fermion. Then, we review the fact that the EMI indeed coincides with a free fermion in $d=2$. Moving to $d\geq 3$, we perform various comparisons between universal coefficients characterizing the EMI and the free fermion theories and show that these differ for both models, the discrepancy growing with $d$.

In section \ref{confblock} we compute the EMI result in the case of two spatially separated and arbitrarily boosted spheres. For a general CFT, such MI can be written as a linear combination of the conformal blocks associated to each primary operator in the replica theory, $I(A,B)=\sum_{\Delta,J} b_{\Delta,J} G^d_{\Delta,J}(u,v)$, where $u,v$ are the relevant conformal cross-ratios. Using the fact that a free fermion controls the leading piece, the EMI is argued to contain a leading piece associated to such field with $\Delta=d-1$ and $J=1$. Interestingly, we find that this is actually the full result for the EMI model, \ie
\begin{equation}
I_{\rm \ssc EMI}(A,B)= b_{d-1,1}^{(\rm \ssc ferm)} G_{d-1,1}^d(u,v)\, .
\end{equation}
In other words, the EMI result for two spheres turns out to coincide with the conformal block associated to the current operator $J_{\mu}=\bar \psi \gamma_{\mu} \psi$. In passing, we obtain some new explicit formulas for $G_{d-1,1}^d(u,v)$ for different values of $d$, some of them not available in the literature.

In section \ref{longdist} we test the hypothesis that the EMI may coincide with the free fermion at long distances for arbitrary regions. Considering rectangular regions for $d=3$ fermions in the lattice, we show that this is not the case. An analytic general-$d$ argument in the same direction is provided using regions with very thin parallel sides.

In section \ref{isemicft} we use the results of the previous sections to argue that the EMI cannot describe the MI of any CFT. This follows from the fact that the free fermion present in the putative CFT describing the EMI decouples ---based on standard QFT observations--- from the rest of the theory. This would imply that the long-distance behavior of the EMI should match the free fermion one for arbitrary regions, which contradicts the results of the previous section.

In section \ref{emilimits} we consider the possibility that the EMI may correspond to some limit of CFTs, something that cannot be discarded from our previous analysis. For instance, if we tried to extract the operator content of the dual holographic theory using the long-distance result for the MI produced by the Ryu-Takayanagi formula ---corresponding to the large-$N$ limit--- we would wrongly conclude that the theory is empty of operators. Considering the structure of the subleading piece produced by the free fermion in the EMI expansion for spheres separated a long distance and the possible ways in which this could be compensated by extra fields, we can prove that ---as opposed to the Ryu-Takayanagi formula--- the EMI cannot correspond to a limit of theories either. 

In section \ref{discu} we make some further comments regarding the interpretation of the EMI model and the possible ways in which the set of known axioms for the MI in QFT may be enhanced.

Appendix \ref{stripsphere} contains a calculation of the entanglement entropy universal coefficients for sphere and strip regions in general dimensions for the EMI model. In appendix \ref{B} we provide a proof of the equivalence between the EMI and the conformal block $G_{d-1,1}^d(u,v)$. Appendix \ref{CB-CC} contains some new formulas for the conformal blocks in the cases corresponding to conserved currents, $\Delta=J+d-2$. Finally, appendix \ref{coeff} includes a calculation of the coefficient appearing in the subleading contribution to the MI of a free fermion for large separations.

% we know an example for which it would fail. the knowledge of the Ryu-Takayanagi formula ---which is the large-$N$ limit result for the entanglement entropy of holographic theories---

%an analysis analogous to the one performed in the previous sections applied to the Ryu-Takayanagi formula would wrongly lead us to the conclusion that the dual theory is empty of operators.  
%the incompatibility of two facts: 1) the EMI must contain a free fermion and therefore its long-distance contribution must equal the one of a free fermion; 2) the coefficient of such contribution ---given that the free fermion must decouple from the rest of the putative EMI theory--- 
 % \comment{blah blah on what we exactly do}

%The remainder of the paper goes as follows

\section{EMI contains a free fermion}\label{long-dist}
In this section, we show that, if the EMI is a CFT in general dimensions, this necessarily contains a free fermion as the lowest dimensional operator. We then review the known fact that in $d=2$ the EMI indeed coincides with the free fermion theory and how this match does not hold in higher dimensions.

The leading term in the expansion of MI for long separating distance $L$ between $A$ and $B$ is known to be dominated by the lowest dimensional primary field (or fields) of the CFT \cite{Cardy.esferaslejanas}. The general expression for this leading term is
\be
I(A,B)\sim C(A,B)\,\left(\frac{R_A R_B}{L^2}\right)^{2 \Delta}\,,  \label{ff}
\ee
where $\Delta$ is the lowest dimension of the theory, $R_A$, $R_B$, are some typical length scales of the regions, and $C(A,B)$ is a dimensionless coefficient. In general, the form of this coefficient depends on the full operator content of the theory. A formula for $C(A,B)$ in terms of the modular flows of the regions $A$ and $B$ has been recently derived in \cite{casini2021mutual}. For the special case where the two entangling surfaces are spheres,  $\partial A=\partial B=\mathbb{S}^{d-2}$, this modular flow is universal and geometric in a CFT \cite{Hislop:1981uh}. As a consequence of that, a closed expression for the coefficient is known in the case of spheres. This was first obtained for scalar fields in \cite{agon2016quantum}. Taking $R_A$ and $R_B$ in (\ref{ff}) as their radii, such coefficient is independent of the spheres orientations in spacetime and reduces to a function of the conformal dimension. We have  \cite{agon2016quantum}
\be \label{ess}
I_{\rm \ssc scal}(A,B)=  c(\Delta)\,\left(\frac{R_A R_B}{L^2}\right)^{2 \Delta}+\dots\,, \quad \text{where} \quad
 c(\Delta)\equiv \frac{\sqrt{\pi} \, \Gamma\!\(2\Delta+1\)}{4\,\Gamma\(2\Delta+\frac32\)}\,.
 \ee
Because the geometric form of the modular flow is independent of the spacetime dimension $d$, this coefficient only depends on $\Delta$.   

The leading contribution of a fermion field was analyzed for spatial spheres in \cite{chen2017mutual}. The case of spheres of arbitrary orientations in spacetime was computed in \cite{casini2021mutual}. Calling $n_A$, $n_B$, to the future directed time-like unit vectors normal to the planes of the spheres, and $l$ to the unit spatial vector in the direction joining the centers of the two spheres, we have for a spinor primary field
\begin{align}
I_{\rm \ssc ferm}(A,B)&=   2^{[\frac{d}{2}]+1} c(\Delta)  \,[2(n_A\cdot l)(n_B\cdot l)-(n_A\cdot n_B)]\, \left(\frac{R_A \, R_B}{L^{2}}\right)^{2\Delta}+\dots
\label{fer}
\end{align}
Interestingly, an explicit calculation for the EMI model  shows that the same tensorial structure appears,
%On the other hand, for the EMI model, the leading long-distance term for spheres has the same tensorial structure 
\be 
I_{\rm \ssc EMI}(A,B)= \frac{4(d-1)(d-2) \pi^{d-1} \kappa_{(d)}}{\Gamma\!\(\frac{d+1}{2}\)^2}\,  \,[2(n_A\cdot l)(n_B\cdot l)-(n_A\cdot n_B)]\, \left(\frac{R_A \, R_B}{L^{2}}\right)^{d-1}+\dots \label{eee}
\ee
This follows from doing the integrals in (\ref{sis2}) and expanding for long-distance. We will be more explicit about this computation in section \ref{confblock} below, where we compute the full form of the EMI mutual information for boosted spheres at any distance. 

Now, the comparison of the EMI expression with \req{ff} shows that the lowest dimensional primary of the EMI model must have dimension $\Delta=(d-1)/2$. Because of the unitarity bounds \cite{Hofman:2008ar,Hofman:2016awc}, this can be only the case of either a scalar field or a fermion field. Comparing with equations (\ref{ess}) and (\ref{fer}) we see that the tensor structure of the contribution is only compatible with a fermion field. This fermion field saturates the unitarity bound and is a free field of helicity $1/2$. We cannot have a contribution of a scalar field of the same dimension as it would spoil the tensor structure in eq. (\ref{eee}).  

In order to compare the free fermion with the EMI, it is useful to calibrate the coefficient $\kappa_{(d)}$ to exactly match the long-distance free fermion contribution by taking   
\be \label{kappad}
\kappa_{(d)}=\frac{  2^{[\frac{d}{2}]}\Gamma\!\(d-2\)\Gamma\!\(\frac{d+1}{2}\)^2 }{8 \pi^{d-3/2} \Gamma\!\(d+\frac{1}{2}\)}\, ,
%1/x\, 2^{[\frac{d}{2}]+1}\frac{\sqrt{\pi} \, \Gamma[d]}{4\,\Gamma[d+1/2]}\,.
\ee
although note that, in principle, the number of free fermionic fields in the EMI could be any integer.

%{\color{blue} Cuanto vale $x$? }

\subsection{Two dimensions}
As it turns out,
in $d=2$ the EMI model indeed coincides with the free fermion. For the latter, in the case of two arbitrary regions formed by intervals, whose projections on the null axis $x^\pm=x^1\pm x^0$ are the multi-interval sets $A^\pm$ and $B^\pm$, we have \cite{casini2005entanglement,casini2009reduced}
\be   
I_{\ssc \rm ferm}(A,B)= \frac{1}{6} \left[\int_{A^+} \diff x^+ \int_{B^+} \diff y^+\, \frac{1}{(x^+-y^+)^2}+\int_{A^-} \diff x^- \int_{B^-} \diff y^-\, \frac{1}{(x^--y^-)^2} \right]\,. 
\ee
This expression is bilinear in the two regions and gives an extensive MI, $I_3\equiv0$. It coincides with \req{aa} for $\kappa_{(2)}=1/6$.

The reason for extensivity in this case can be tracked to bosonization in $d=2$ \cite{casini2005entanglement}. The R\'enyi operators in the replicated theory are products of exponentials of the current for $n$ different decoupled free fermion fields. The free fermion current can be written as a linear expression in a dual free scalar field. Therefore the correlators of the R\'enyi twist operators are correlators of exponentials of free fields. These are exponentials of bilinear expressions, and upon taking the logarithm, the R\'enyi entropies turn out to be bilinear in the two regions. The entropy inherits this same form.

\subsection{Not a free fermion for $d>2$}
\label{not}
In dimensions higher than two, the free fermion does not have extensive mutual information. This can be seen in several ways by comparing the EMI results for various entangling regions with the analogous free fermion ones. In this subsection, we perform several comparisons of that kind and observe that the discrepancy in various charges characterizing the free fermion and the EMI tends to grow as the spacetime dimension increases. For many of the comparisons, we normalize the EMI results so that the long-distance coefficient matches the free fermion one, as in \req{kappad}. We also try with other ratios. 

%\subsubsection{Three dimensions}
{\bf Three dimensions.} Let us start with the $d=2+1$ case. Consider first the mutual information for two regions with parallel boundaries of size $L$ separated a short distance $r\ll L$. As reviewed in appendix \ref{stripsphere}, this behaves, for any CFT as
\begin{equation}\label{strip}
I= \frac{k^{(3)} L}{r} + \dots
\end{equation} 
where $k^{(3)}$ is a theory-dependent coefficient which matches the universal coefficient in the entanglement entropy corresponding to a long and thin strip of dimensions $L\times r$. For the EMI model, the result for the general-dimension version of $k^{(d)}$ appears in \req{kemid}. Particularizing to $d=3$, we have
\begin{equation}
k^{(3)}_{\rm \ssc EMI}=2\pi \kappa_{(3)} \overset{ \ssc (\ref{kappad})}{=} \frac{4}{15\pi}\simeq 0.0849 \, ,
\end{equation}
where in the second equality we fixed the value of $\kappa_{(3)}$ calibrated so that the long-distance coefficient for a pair of disks matches the free fermion one ---see \req{kappad} above.
For a Dirac field, the analogous coefficient reads \cite{Casini:2009sr}
\begin{equation}
k^{(3)}_{\rm \ssc ferm}\simeq 0.0722\, .
\end{equation}
Hence, we observe that both values are clearly different, the discrepancy being $\sim 15\%$.

%We can use this to fix $\kappa_{(3)}\simeq 0.02298$.
On the other hand, if one considers the mutual information for two concentric disks of radii $R_A<R_B$, the EMI result obtained in the following section for $d=3$ becomes
\begin{equation}
I_{\rm  \ssc EMI}=8\pi^2 \kappa_{(3)} \frac{(R_A/R_B)^2}{[1-(R_A/R_B)^2]}\, .
\end{equation}
As explained in the appendix, when the disks are very close to each other, $R_A=R-\delta/2$, $R_B=R+\delta/2$, $\delta\ll R$, we can expand this expression to get
\begin{equation}
I_{\rm  \ssc EMI}=k^{(3)}_{\rm \ssc EMI} \frac{2\pi R}{\delta }- 2 F^{(3)}_{\rm \ssc EMI}\, , \quad \text{where}\quad F^{(3)}_{\rm \ssc EMI}=2\pi^2 \kappa_{(3)}\overset{ \ssc (\ref{kappad})}{=}\frac{4}{15}\simeq 0.2667  \, ,
\end{equation}
is the $F$-term appearing in the entanglement entropy of a disk region, and the strip coefficient $k^{(3)}_{\rm \ssc EMI}$ appears in the area-law piece.   In the case of the Dirac fermion, $F^{(3)}$ is known analytically and reads \cite{Klebanov:2011gs,Marino:2011nm}
\begin{equation}
F^{(3)}_{\rm \ssc ferm} =\frac{1}{8} \left(2 \log 2+ \frac{3}{\pi^2} \zeta(3) \right) \simeq 0.21896\, . 
\end{equation}
Hence, we again observe a considerable difference between models, in this case around $\sim 18\%$. 

Observe that by using the value of $\kappa_{(3)}$ calibrated with the long distance coefficient for different charges such as $k^{(3)}$ or $F^{(3)}$, what we are effectively doing is comparing the quotient of such quantities divided by the long distance coefficient in both models. We can naturally perform additional comparisons by dividing by other charges. For instance,  comparing the quotients $F^{(3)}/k^{(3)}$ instead, one obtains
\begin{equation}
\frac{F^{(3)}_{\rm \ssc EMI}}{k^{(3)}_{\rm \ssc EMI}}=\pi \, , \quad \frac{F^{(3)}_{\rm \ssc ferm}}{k^{(3)}_{\rm \ssc ferm}} \simeq 3.033 \, ,
\end{equation}
which are much closer, only differing by $\sim 3.5\%$.

Additional comparisons can be made by considering an entangling region with a corner of angle $\theta$. In that case, the entanglement entropy contains a logarithmic divergence of the form \cite{Casini:2006hu,Hirata:2006jx}
\begin{equation}
\see|_{\log}= -a(\theta) \log (R/\delta)\, ,
\end{equation}
where $a(\theta) $ is a universal function of the corner opening angle. In the case of the EMI, this function turns out to be given by \cite{Casini:2008wt,Swingle:2010jz}
\begin{equation}\label{emicor}
a_{\rm \ssc EMI}(\theta)=2\kappa_{(3)} \left[1+(\pi -\theta) \cot \theta \right] \, .
\end{equation}
For very sharp and almost-smooth corners, $a(\theta)$ behaves, on general grounds, as \cite{Casini:2006hu}
\begin{equation}
a(\theta \rightarrow 0) =\frac{k^{(3)} }{\theta}+ \dots\, , \quad a(\theta \rightarrow \pi)= \sigma (\pi -\theta)^2+ \dots
\end{equation}
In this expression, $k^{(3)}$ is the strip coefficient appearing in \req{strip}, and $\sigma$ turns out to be related to the stress-tensor two-point function charge $\ctt$\footnote{For a general CFT in $d$ dimensions, the tensorial structure of the flat-space stress tensor two-point function is fully determined by conformal symmetry up to a theory-dependent constant:  $\braket{T_{ab}(x)T_{cd}(0)}_{\mathbb{R}^d}=\ctt \mathcal{I}_{ab,cd}/|x|^{2d}$, where $\mathcal{I}_{ab,cd}$ is a fixed dimensionless expression \cite{Osborn:1993cr}.} as \cite{Bueno:2015rda,Faulkner:2015csl}
\begin{equation}
\sigma=\frac{\pi^2}{24} \ctt\, .
\end{equation}
Using \req{emicor}, we can then obtain the would-be value of $\ctt^{\rm \ssc EMI}$ in case this represented an actual QFT. The result, and the analogous one for a free fermion \cite{Osborn:1993cr} are given respectively by
\begin{equation}
\ctt^{\rm \ssc EMI}=\frac{16 \kappa_{(3)}}{\pi^2}\overset{ \ssc (\ref{kappad})}{=}\frac{32}{15\pi^2}\simeq 0.0219\, , \quad \ctt^{\rm \ssc ferm}=\frac{3}{16\pi^2}\simeq 0.019\, .
\end{equation}
These two differ by $\sim 13\%$. We can also compare the results with $F^{(3)}$ or $k^{(3)}$. For instance, we have
\begin{equation}
\frac{F^{(3)}_{\rm \ssc EMI}}{\ctt^{\rm \ssc EMI}}=\frac{\pi^4}{8}\simeq 12.1761  \, , \quad \frac{F^{(3)}_{\rm \ssc ferm}}{\ctt^{\rm \ssc ferm}}=\frac{\pi^2}{3} \left( 2 \log 2 + \frac{3 \zeta(3)}{\pi^2} \right)\simeq 11.5256  \, ,
\end{equation}
which differ by $\sim 5.3\%$. We observe that comparing the various charges with the long-distance coefficient tends to make the agreement with the fermion worse than when comparing the other charges amongst each other.

%\subsubsection{Four dimensions}
{\bf Four dimensions.} In this case we can use Solodukhin's formula for the entanglement entropy universal term to evaluate the trace-anomaly coefficients $a$ and $c$  \cite{Solodukhin:2008dh}. In particular for entangling surfaces corresponding to spheres and cylinders the relevant terms read, respectively,
\begin{equation} \label{ddd}
\see |_{\rm \ssc sphere}=-4 a \log (R/\delta)  \, , \quad \see |_{\rm \ssc cylinder}= -\frac{c}{2 }  \frac{L}{R} \log (R/\delta)\, .
\end{equation}
For the EMI, simple calculations yield the right expressions appearing in \req{ddd}, where
\begin{equation}
a_{\rm \ssc EMI}= \pi^2 \kappa_{(4)}\overset{ \ssc (\ref{kappad})}{=} \frac{3}{70}\simeq 0.0429\, , \quad c_{\rm \ssc EMI}=\frac{3\pi^2 \kappa_{(4)}}{2}\overset{ \ssc (\ref{kappad})}{=}\frac{9}{140}\simeq 0.0643\, , \quad \text{so} \quad \frac{   a_{\rm \ssc EMI}}{c_{\rm \ssc EMI}} = \frac{2}{3} \, .
\end{equation}
It is an interesting fact that the would-be CFT represented by the EMI lies within the range of values allowed by the unitarity bounds \cite{Hofman:2008ar,Hofman:2016awc} 
\begin{equation}
\frac{1}{3} \leq \frac{a}{c} \leq \frac{31}{18}\, .
\end{equation}
Naturally, the free fermion also satisfies the bounds, but one finds in that case \cite{Hofman:2008ar,Dowker:1976zf,Duff:1977ay}
\begin{equation}
a_{\rm \ssc ferm}=\frac{11}{360}\simeq 0.0306\, \quad c_{\rm \ssc ferm}=\frac{1}{20}=0.05\, \quad \frac{   a_{\rm \ssc ferm}}{c_{\rm \ssc ferm}} =\frac{11}{18} \simeq 0.6111\, .
\end{equation}
The discrepancies range from $\sim 8.33\%$ for the ratios $a/c$ to $\sim 29\%$ for the $a$'s normalized by the long-distance coefficient.

 We can make another comparison using the values of the coefficient characterizing the mutual information of two parallel regions which are very close to each other ---see appendix \ref{stripsphere}.  %Indeed, the $d$-dimensional generalized version of \req{strip} reads \cite{Casini:2009sr}
% \begin{equation}
 %I=k^{(d)}  \frac{\mathcal{A}}{r^{d-2}}+\dots
 %\end{equation}
 %where $r$ is the separation between the faces of $A$ and $B$, and $\mathcal{A}$ stands for the area of the parallel faces. 
 The result for $k^{(4)}$ for a Dirac fermion is given by \cite{Casini:2009sr}
 \begin{equation}
 k^{(4)}_{\rm \ssc ferm} \simeq 0.0215\, .
 \end{equation}
  whereas for the EMI we find from \req{kemid}
  \begin{equation}
 k^{(4)}_{\rm \ssc EMI} = 2\pi \kappa_{(4)}\overset{ \ssc (\ref{kappad})}{=} \frac{3}{35\pi}\simeq 0.0273\, ,
  \end{equation} 
 which differ by $\sim 21\%$.  On the other hand, the ratios $a/k^{(4)}$ are a bit closer, namely
   \begin{equation}
\frac{a_{\rm \ssc EMI}}{k^{(4)}_{\rm \ssc EMI}}=\frac{\pi}{2}\simeq 1.5708 \, , \quad \frac{a_{\rm \ssc ferm}}{k^{(4)}_{\rm \ssc ferm}} \simeq 1.4199  \, ,
\end{equation}
  which is $\sim 9.6\%$ off.

% \subsubsection{Five dimensions}
{\bf Five dimensions.} Moving to $d=4+1$, we can consider for instance the strip coefficient $k^{(5)}$ and the one appearing in the entanglement entropy across a $\mathbb{S}^3$, which we denote $F^{(5)}$. As explained in appendix \ref{stripsphere}, both can be obtained from the mutual information of a pair of concentric spheres of radii $R_A=R-\delta/2$, $R_B=R+\delta/2$, $\delta \ll R$, analogously to lower dimensions. In this case we have, on general grounds
 \begin{equation}
 I= k^{(5)} \frac{2\pi^2 R^3}{\delta^3}+ \dots+ 2 F^{(5)}\, ,
 \end{equation}
 where the dots denote subleading divergences. For the EMI, we find
 \begin{equation}
 k^{(5)}_{\rm \ssc EMI}=\frac{\pi^2 \kappa_{(5)}}{2}\overset{ \ssc (\ref{kappad})}{=}\frac{64}{945\pi^2}\simeq 0.00686\, , \quad F^{(5)}_{\rm \ssc EMI}=\pi^4 \kappa_{(5)}\overset{ \ssc (\ref{kappad})}{=}\frac{128}{945}\simeq 0.135\, , 
 \end{equation} 
 so
 \begin{equation}
  \frac{F^{(5)}_{\rm \ssc EMI} }{k^{(5)}_{\rm \ssc EMI}}=2 \pi^2 \simeq 19.7392\, .
 \end{equation}
 On the other hand, for the fermion, we have \cite{Casini:2009sr,Klebanov:2011gs}
  \begin{equation}
 k^{(5)}_{\rm \ssc ferm}=0.0052  \, , \quad F^{(5)}_{\rm \ssc ferm }=\frac{1}{64} \left[6\log 2 + \frac{10 \zeta(3)}{\pi^2}+\frac{15 \zeta(5)}{\pi^4} \right] \simeq 0.0865 \, ,  \quad \text{so} \quad \frac{F^{(5)}_{\rm \ssc ferm } }{k^{(5)}_{\rm \ssc ferm }} \simeq 16.636 \, .
 \end{equation} 
 Hence, in this case the differences range from $\sim 15.7\%$ for the quotient $F^{(5)}_{\rm \ssc ferm }/k^{(5)}_{\rm \ssc ferm }$ to $\sim 35.9 \%$ for the  $F^{(5)}_{\rm \ssc ferm }$ normalized by the long-distance coefficient.
 % \comment{add also strip result for EMI in $d=4$}
 
 % \subsubsection{Six dimensions}
{\bf Six dimensions.} In $d=5+1$ we can consider again the coefficient $A^{(6)}$ which appears weighting the Euler density in the trace anomaly ---see \req{traceano} below. This we can compare with the strip one, $k^{(6)}$. From \req{kemid} and \req{FAemi} we have for the EMI model
 \begin{equation}
 k_{\ssc \rm EMI}^{(6)}=\frac{\pi^2}{3}\kappa_{(6)}\overset{ \ssc (\ref{kappad})}{=}\frac{10}{231\pi^2}\simeq 0.00439 \, , \quad
 A_{\ssc \rm EMI}^{(6)}=\frac{\pi^4}{6}\kappa_{(6)}\overset{ \ssc (\ref{kappad})}{=} \frac{5}{231}\simeq 0.0216 \, ,
 \end{equation} 
 so
 \begin{equation}
 \frac{A_{\ssc \rm EMI}^{(6)}}{k_{\ssc \rm EMI}^{(6)}}=\frac{\pi^2}{2}\simeq 4.9348\, .
 \end{equation}
 On the other hand, the analogous coefficients for the free fermion can be extracted from refs. \cite{Casini:2009sr} and \cite{Bastianelli:2000hi,Safdi:2012sn}, respectively, and read  
  \begin{equation}
 k_{\ssc \rm ferm}^{(6)}\simeq 0.00325 \, , \quad
 A_{\ssc \rm ferm}^{(6)}=\frac{191}{15120}\simeq 0.0126  \quad \text{so} \quad \frac{A_{\ssc \rm ferm}^{(6)}}{k_{\ssc \rm ferm}^{(6)}}\simeq 3.889 \, .
 \end{equation} 
 In this case, discrepancies vary from $\sim 26\%$ for the $k^{(6)}$'s normalized by the long distance coefficient to $\sim 41.7\%$ for the $A^{(6)}$'s normalized by the same.
 
{\bf Seven dimensions.} As a final case, we consider $d=6+1$. Once more, we compare $k^{(7)}$ and $F^{(7)}$. Using the formulas in the appendix, we have for the EMI
 \begin{equation}
 k_{\ssc \rm EMI}^{(7)}=\frac{\pi^3}{16}\kappa_{(7)} \overset{ \ssc (\ref{kappad})}{=} \frac{256}{5005\pi^3}\simeq 0.00165 \, , \quad
F_{\ssc \rm EMI}^{(7)}=\frac{\pi^6}{12}\kappa_{(7)} \overset{ \ssc (\ref{kappad})}{=} \frac{1024}{15015} \simeq 0.0681 \, .
 \end{equation} 
 so
 \begin{equation}
 \frac{F_{\ssc \rm EMI}^{(7)}}{k_{\ssc \rm EMI}^{(7)}}=\frac{4\pi^3}{3}\simeq 41.342\, .
 \end{equation}
 On the other hand, for a free fermion we have \cite{Casini:2009sr,Klebanov:2011gs}
   \begin{equation}
 k_{\ssc \rm ferm}^{(7)}\simeq  0.0012\, , \quad
F_{\ssc \rm ferm}^{(7)}=\frac{1}{512} \left[20 \log 2 + \frac{518 \zeta(3)}{15\pi^2}+\frac{70 \zeta(5)}{\pi^4}+ \frac{63 \zeta(7)}{\pi^6} \right]  \simeq 0.0369   \, .
 \end{equation} 
 so 
 \begin{equation}
 \frac{F_{\ssc \rm ferm}^{(7)}}{k_{\ssc \rm ferm}^{(7)}}\simeq 30.73\, .
 \end{equation}
The greatest difference ($\sim 45.8\%$) appears in the $F^{(7)}$'s normalized by the long distance coefficient.

 All in all, we observe that both models are clearly different for $d>2$ and they seem to increasingly deviate from each other as the number of dimensions  grows.

\section{Current conformal block equals EMI for spherical regions}\label{confblock}
In this section we compute the MI  for two spatially separated and arbitrarily boosted spheres for the EMI model. Remarkably, the result turns out to be identical to the one corresponding to the conformal block $G_{d-1,1}^d$ associated to the conserved current operator made out of two free fermions.

In Minkowski space, the configuration space of ``boosted'' spherical regions is determined by the 
tips of their causal developments. Thus, for two spheres, the mutual information is a function of the four associated spacetime points. Furthermore, conformal symmetry implies that such dependence can only come through two independent conformally invariant cross ratios, say,  $u$ and $v$ ---see below for the precise definitions.

In the long-distance regime, an OPE analysis in the replicated theory suggests that such function should be given in terms of the conformal blocks associated to each primary operator of the replica theory. The following expansion is implied \cite{Long:2016vkg, Chen:2016mya, Chen:2017hbk},
\bea\label{I-expansion}
I(A,B)=\sum_{\Delta,J}b_{\Delta,J} G^d_{\Delta,J}(u,v)\, .
\eea
Here, $\Delta$ and $J$ are the scaling dimension and spin of the primary modules in the replica theory,\footnote{Notice that it might be the case that the operator $\mathcal{O}_\Delta$ is a primary operator in the replica theory but it is not a primary in the original CFT.} respectively, and the coefficients $b_{\Delta,J}$ of each contributing conformal block can be obtained via an analysis along the lines of \cite{Cardy.esferaslejanas,agon2016quantum} ---this was recently reviewed, generalized and expressed in a compact form in \cite{casini2021mutual}.  In particular, for spheres lying on the same time slice, the sum of the coefficients associated to the leading contribution of a primary operator in the original CFT takes the simple form
\bea
\sum_J b_{\Delta,J}=\frac{\sqrt{\pi} \,\,  \Gamma\!\(\Delta +1\){\rm dim}(\mathcal{R})}{4^{\Delta+1}\Gamma\!\(\Delta+\frac32\)}\,,
\eea
where $\mathcal{R}$ is the Lorentz representation of the CFT primary operator that contributes to $I(A,B)$. For other primary operators in the replica theory, one can either use the analysis of \cite{casini2021mutual} adapted to such computation or the explicit formulas from \cite{Long:2016vkg, Chen:2016mya, Chen:2017hbk}.

Now, if we are able to compute the mutual information for boosted spherical regions in the EMI model, equating such result to the RHS of \req{I-expansion} would allow us to extract the full set of $\{\Delta,J\}$ of the primary modules contributing to the mutual information of the putative CFT corresponding to this model. 
From the analysis of section \ref{long-dist} we know that the EMI is dominated at long distances by the contribution coming from a free fermion. Comparing that result with \req{I-expansion}, we find that the leading replica operator that contributes to $I_{\rm \ssc EMI}(A,B)$ ---through analytic extension from the contribution to the Renyi entropies to replica number $n=1$--- is precisely a conserved current operator made out of two free fermions in two replicas $i,j$, $i\neq j$,
\bea
J_\mu =\bar{\psi}_i\, \gamma_\mu \psi_j\,,
\eea
with $\Delta=d-1$ and $J=1$. Taking this as our starting point, we would like to find the set of operators that contribute to $I_{\rm \ssc EMI}(A,B)$ via a recursive process in which we consecutively subtract the various conformal-block contributions to the full expression of the EMI.

In more detail, the idea is the following. We start assuming that $I_{\rm \ssc EMI}(A,B)$ has an expansion as the one given in \req{I-expansion}, thus 
\bea\label{I-emi}
I_{\rm \ssc EMI}(A,B)=\sum_{\Delta,J}b_{\Delta,J} G^d_{\Delta,J}(u,v)\, ,
\eea
for a set $\{\Delta, J\}$ to be determined. Since at long distances the leading contribution is consistent with the conformal block of an operator with $\Delta=d-1$ and $J=1$, and such operator can only be associated to the leading contribution of a free fermion, we subtract such contribution to $I_{\rm \ssc EMI}(A,B)$ and repeat the analysis for the remaining expression. This is, after the first iteration we have 
\bea\label{first-ite}
I^{(1)}_{\rm \ssc EMI}(A,B)\equiv I_{\rm \ssc EMI}(A,B)-b^{\rm \ssc (ferm)}_{d-1,1}G^d_{d-1,1}(u,v)
\eea
with \cite{Chen:2017hbk}
\bea\label{bf}
b^{\rm \ssc (ferm)}_{d-1,1}=2^{[\frac{d}2]+1}\frac{\sqrt{\pi}\, \Gamma\!\(d\)}{4^d\, \Gamma\!\(d+\frac{1}{2}\)}\, ,
\eea
which has the expansion
\bea
I^{(1)}_{\rm \ssc EMI}(A,B)=\sum_{\{\Delta,J\}\neq\{d-1,1\} }b_{\Delta,J} G^d_{\Delta,J}(u,v)\,.
\eea
Studying the long-distance behavior of (\ref{first-ite}) allows us to deduce the associated values of $\{\Delta, J\}$ for the next-to-leading contributing operator and by reading the associated coefficient we could deduce to which operator that corresponds to. We repeat the process until we have exhausted all the operators that contribute to the LHS of (\ref{I-emi}) or until we have zero remanent. The process will turn out to be surprisingly short, as we will find $I^{(1)}_{\rm \ssc EMI}(A,B)=0$.

\subsection{EMI for boosted spherical regions}

\begin{figure}
\begin{center}
\includegraphics[scale=0.4]{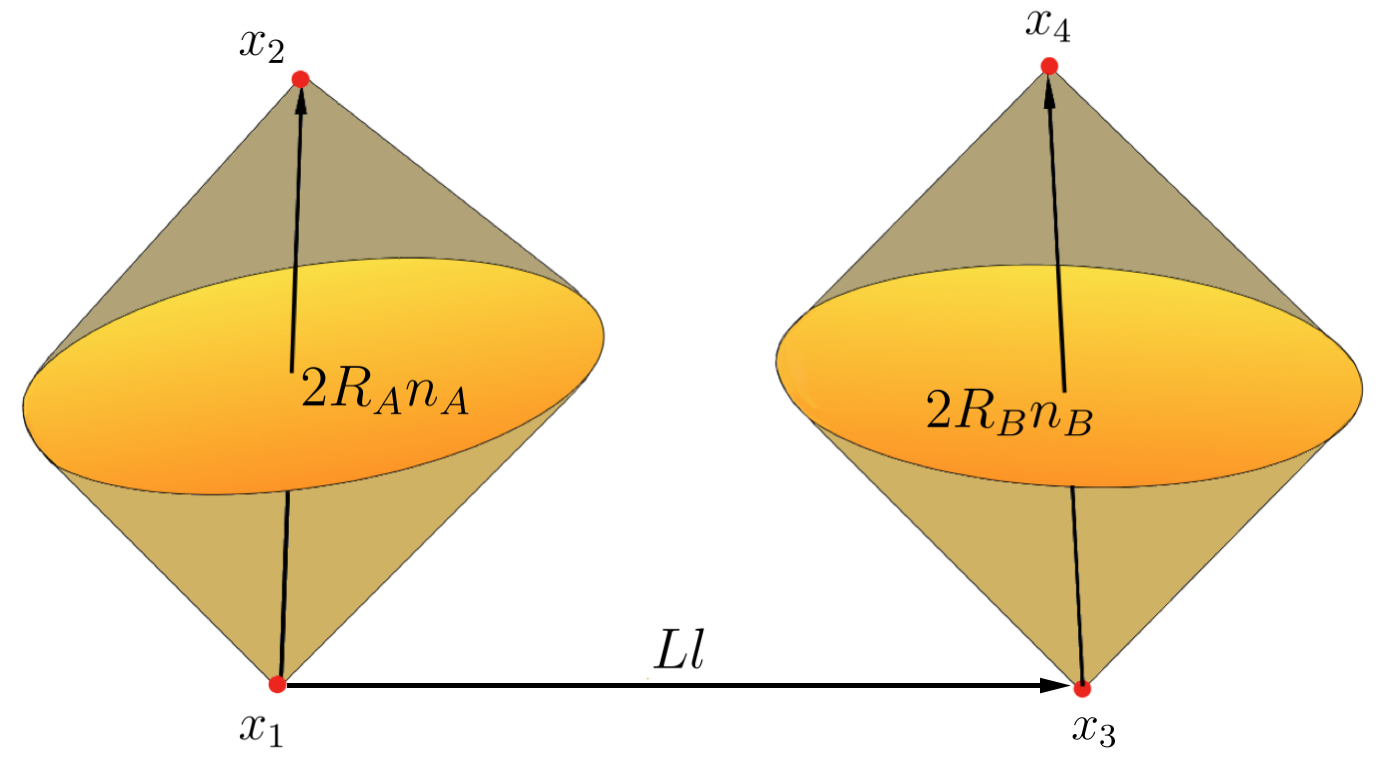}
\end{center}
\caption{Schematic representation of the geometric setup. We consider two boosted spheres $\mathbb{S}_A^{d-2}$, $\mathbb{S}^{d-2}_B$, characterized by their causal tips $x_1$,$x_2$ and $x_3$,$x_4$, respectively. The vector joining $x_1$ and $x_3$ is denoted $L l$ and the ones connecting the diamond tips of each sphere by $2R_A n_A$ and $2R_B n_B$, respectively.}
\label{fig:disj-spheres}
\end{figure}

The first step in our program is to compute the EMI for relatively boosted spherical regions. In such case the EMI formula is given by \req{sis2} or, equivalently, by \req{sis2}. %\cite{Casini:2008wt}
%\bea
%I_{\rm \ssc EMI}(A,B)&=&\kappa_{(d)}  \int_{\partial A} d\sigma_A^{\alpha \beta} \int_{\partial B} d\sigma^{B}_{\alpha \beta} \, \frac{1}{|{\bf r}_1 - {\bf r}_2|^{2(d-2)}}\,,
%\eea
%where $d\sigma_A^{\alpha \beta}$ is the Hodge dual to the co-dimension two area form induced on $\partial A$. $\bf{r}_1$ and $\bf{r}_2$ denote the points on $\partial A$ and $\partial B$ respectively. In the case of a spherical region we can chose the basis for the index structure to be given by the mutually orthogonal normal vectors $n^\alpha$ and $\bar{n}^\beta$, where $n^\alpha$ is the vector normal to the hyperplane on which the sphere is located and $\bar{n}^\beta$ is a normal vector to the sphere on that hyperplane. Thus 
%\bea
%d\sigma_A^{\alpha \beta}=\(n_A^\alpha \bar{n}_1^\beta-n_A^\beta \bar{n}_1^\alpha \)d\sigma_A
%\eea
%and similarly for $d\sigma^B_{\alpha \beta}$. 
Thus for the case of relatively boosted spheres the mutual information can be written as 
\bea\label{EMI-spheres}
I_{\rm \ssc EMI}(A,B)&=&2\kappa_{(d)}  \int_{\mathbb{S}_A^{d-2}} \diff \sigma_A \int_{\mathbb{S}_B^{d-2}} \diff \sigma_B\, \frac{(n_A\cdot n_B)( \bar{n}_A\cdot \bar{n}_B)-(n_A\cdot \bar{n}_B)( \bar{n}_A\cdot n_B) }{|{\bf r}_A - {\bf r}_B |^{2(d-2)}}\, ,
\eea
where we modified the notation slightly with respect to \req{sis2} for future convenience.
We describe our geometric setup through the tips of the causal diamonds whose past and future null cones intersect at the corresponding spheres. We use $x_1$ and $x_2$ to label the past and future tips of the sphere $A$, respectively, and analogously with  $x_3$ and $x_4$ for the sphere $B$. In figure \ref{fig:disj-spheres}, we give a schematic representation of this setup. 

Given these four spacetime points, the mutual information will be a function of the usual conformal invariant cross ratios $u$, $v$, which are given by
\bea
u\equiv \frac{|x_1-x_2|^2|x_3-x_4|^2}{|x_1-x_3|^2|x_2-x_4|^2}, \qquad {\rm and}\qquad v\equiv\frac{|x_1-x_4|^2|x_2-x_3|^2}{|x_1-x_3|^2|x_2-x_4|^2}\,.
\eea
However, for our convenience we will work instead with the conformally independent parameters 
\bea
\chi_1=\frac{|x_1-x_2||x_3-x_4|}{|x_1-x_3||x_2-x_4|} \quad {\rm and}\quad
\chi_2=\frac{|x_1-x_2||x_3-x_4|}{|x_1-x_4||x_2-x_3|}\,,
\eea
which are related to $u$ and $v$ via 
\bea\label{uv}
u=\chi_1^2, \qquad {\rm and}\qquad v=\frac{\chi_1^2}{\chi_2^2}\,.
\eea
For the above geometric set up we can write 
$
x_2-x_1=2R_A n_A$,  $x_4-x_3=2R_B n_B
$,
where $R_A$ and $R_B$ are the radii of the spheres $A$ and $B$ respectively, and  $n_A$ and $n_B$ are the future directed unit time-like vectors pointing from the past to the future cone tips of each sphere. We write the difference 
$
x_3-x_1=Ll
$, 
where $L$ is the distance between the past tip cones of the spheres and $l$ is a unit space-like vector. 

With the above identifications the conformal parameters become
\bea\label{chi1chi2-disconc}
\chi_1= \frac{4R_AR_B}{L|2R_An_A-Ll-2R_Bn_B|}\,, \quad {\rm and}\quad 
\chi_2=\frac{4R_AR_B}{|2R_An_A-Ll||2R_Bn_B+Ll|}\, ,
\eea
with ranges $\chi_1\in \(0,1\)$, $\chi_2\in \(0,\infty\)$
with $\chi_2 >\chi_1$ for spatially separated spheres. 

Although we are interested in the above geometric set up, we can simplify our computation, without loss of generality, by performing a conformal transformation which makes the centers of the spheres coincide.  In that case, the conformal parameters take the values  
\bea\label{chi1chi2-conc}
\chi_1=\frac{4R_AR_B}{|R_An_A-R_Bn_B|^2}\,, \qquad {\rm and}\qquad
\chi_2=\frac{4R_AR_B}{|R_An_A+R_Bn_B|^2}\,,
\eea
where we use the relation
$
x_3-x_1=R_An_A-R_Bn_B=Ll
$
that follows from the equal-center condition. We parametrize the inner product between $n_A$ and $n_B$ by a boost parameter $\beta$ via 
\bea \label{boost}
n_A\cdot n_B\equiv -\cosh\beta\,,
\eea
and introduce the parameter $0\leq x\leq 1$, defined as
\bea\label{x}
x\equiv \frac{2R_AR_B}{R_A^2+R_B^2}.
\eea
The parameters $\chi_1$ and $\chi_2$ can then be written in terms of $x$ and $\beta$ as
\bea\label{chi1chi2-xbeta}
\chi_1=\frac{2x}{|1-x\cosh\beta|}\,, \quad {\rm and}\quad 
\chi_2=\frac{2x}{\(1+x\cosh\beta\)}\,.
\eea
In this case, $\chi_1>\chi_2$ and $\chi_2\in \(0,1\)$, $\chi_1\in \(0,\infty\)$ so essentially the conformal transformation interchanged the r\^oles of $\chi_1$ and $\chi_2$, which is a natural thing for an inversion like the one we are considering here. %transformation as the one required to go from the disjoint to the ``concentric'' case. 

Given $R_A$ and $R_B$, there is a critical value of $\beta$ for which $\chi_1$ is ill-defined. From \req{chi1chi2-conc} one can see that this happens when the tip cones are null-like separated, 
$
R_An_A-R_Bn_B={\rm null \,\, vector}
$, 
or in other words, when the causal cones have a common boundary. Assuming the casual cones obey a strict inclusion relation, then \bea\label{beta}
0\leq \cosh\beta <\frac{R_A^2+R_B^2}{2R_A R_B}\,.
\eea
This in turn is equivalent to say that the cone tips are time-like separated. To describe such configurations, the absolute value in (\ref{chi1chi2-xbeta}) is thus unnecessary.  

We will write down a formula for the EMI in the above set up as a funciton of $x$ and $\beta$. Then, whenever necessary, we will go back and rewrite it in terms of the perhaps more familiar $u$ and $v$ variables. Points on the spheres $A$ and $B$ will be denoted by ${\bf r}_A$ and ${\bf r}_B$ respectively. $n_A$ and $n_B$ are also the future oriented normal vectors to the spheres $A$ and $B$ respectively. Thus the points on the spheres satisfy
\bea\label{spheres}
{\bf r}_A^2=R_A^2, \quad {\bf r}_A^2=R_A^2, \qquad {\rm and} \qquad n_A\cdot {\bf r}_A=0\,, \quad n_B\cdot {\bf r}_B=0\,.
\eea
 Since the spheres are concentric, we can write ${\bf r}_A=R_A \bar{n}_A$ and ${\bf r}_B=R_B\bar{n}_B$ where $\bar{n}_A$, $\bar{n}_B$ are the unit vectors normal to each sphere on their respective hyperplanes.

We can choose a coordinate system such that the sphere $A$ lies on the $t=$ constant slice while the sphere $B$ is boosted along the $\hat{z}$ direction with the respect to $A$. We have then
\bea\label{normals}
n_A=\hat{t}, \qquad {\rm and } \qquad n_B=\cosh\beta \hat{t}+\sinh\beta \hat{z}\,.
\eea
Given eq. (\ref{normals}), we can find the solutions to eq. (\ref{spheres}) as 
\bea \label{ra}
&&{\bf r}_A=R_A \bar{n}_A
=R_A\(|\vec{\Omega}_{A}|\hat{\Omega}_{A}+\Omega_{A,z}\hat{z} \)\,, \\
&&{\bf r}_B=R_B \bar{n}_B=R_B\(\sinh\beta \Omega_{B,z} \hat{t}+|\vec{\Omega}_B|\hat{\Omega}_B+ \cosh \beta \Omega_{B,z}\hat{z}\)\,.
\eea
The relative boost partially breaks the spherical symmetry and thus we have separated the angular variables describing a unit sphere by $|\vec{\Omega}|\hat{\Omega}+\Omega_z\hat{z}$ above. We will integrate first the sphere $A$ which is the most symmetric one in the above coordinates. To simplify our computation even further, we rotate the coordinate system such that the new $\hat{z}$ coincides with the radial direction of the sphere $B$. This means that the vectors transform as 
\bea
&&{\bf r}_B=R_B\(\sinh\beta \cos\theta_B\hat{t}+
\sqrt{1+\sinh^2 \beta \cos^2\theta_B} \hat{z}\)\,, \nonumber \\
&&n_B=\(\cosh\beta\hat{t}-\sinh\beta \sin\phi_B \,\hat{\Omega}_B+\sinh\beta \cos\phi_B \hat{z}\)\,,
\eea
where we used $\Omega_{B,z}=\cos\theta_B$, and the rotation angle $\phi_B$ can be seen, from the orthogonality condition (\ref{spheres}), to be given by 
\bea
\cos\phi_B=\frac{\cosh\beta \cos\theta_B}{\sqrt{1+\sinh^2\beta \cos^2\theta_B}}\,\quad {\rm and} \quad \sin\phi_B=\frac{\sin\theta_B}{\sqrt{1+\sinh^2\beta \cos^2\theta_B}}\,.
\eea
Since the sphere $A$ is spherically symmetric, we can use expressions (\ref{ra}) and (\ref{spheres}) for ${\bf r}_A$ and $n_A$. 
%it can be put in the same form as before, this is 
%\bea
%{\bf r}_A=R_A\(\vec{\Omega}'+\Omega'_z\hat{z} \)\,, \qquad {\rm and}\qquad n_A=\hat{t}\,.
%\eea
We are ready now to write down the integral expressing the EMI for two spheres using (\ref{EMI-spheres}). The main ingredients are 
 \bea
 &&|{\bf r}_A - {\bf r}_B |^2=\Big|R_A^2+R_B^2-2R_AR_B\sqrt{1+\sinh^2\beta \cos^2\theta_B}\cos\theta_A\Big|\,, \\ \label{numerator}
&& (n_A\cdot n_B)( \bar{n}_A\cdot \bar{n}_B)-(n_A\cdot \bar{n}_B)( \bar{n}_A\cdot n_B)=-\frac{\cosh\beta \cos\theta_A}{\sqrt{1+\sinh^2\beta \cos^2\theta_B}}\,,
 \eea
where in equation (\ref{numerator}) we have ignored terms proportional to $\hat{\Omega}_B\cdot \hat{\Omega}_A$ since those integrate to zero. Likewise, we can explicitly integrate the angles $\hat{\Omega}_B$ and $\hat{\Omega}_A$ for each sphere which entails the replacement
\bea
\int_{\mathbb{S}_A^{d-2}}  \diff \sigma_A \int_{\mathbb{S}_B^{d-2}}  \diff \sigma_B\to \frac{4R_A^{d-2}R_B^{d-2}\pi^{d-2}}{\Gamma\!\(\frac{d-2}{2}\)^2}\int_0^\pi\, \diff \theta_A  \sin^{d-3}\theta_A \int_0^\pi\, \diff \theta_B \sin^{d-3}\theta_B \,.
\eea
Putting everything together, we finally have
\bea
I_{\rm \ssc EMI}&=&\frac{8 \kappa_{(d)}R_A^{d-2}R_B^{d-2}\pi^{d-2}}{\Gamma\!\(\frac{d-2}{2}\)^2}\int_0^\pi \! \! \int_0^\pi \frac{\diff \theta_B  \diff \theta_A\, \sin^{d-3}\theta_B \, \sin^{d-3}\theta_A}{\Big|R_A^2+R_B^2-2R_AR_B\sqrt{1+\sinh^2\beta \cos^2\theta_B}\cos\theta_A \Big|^{d-2}}\, \nonumber \\
&&\qquad \qquad  \qquad \qquad \qquad \qquad \times \frac{\cosh\beta \cos\theta_A}{\sqrt{1+\sinh^2\beta \cos^2\theta_B}}\,.
\eea
We can rewrite the above integral in terms of the parameter $x$ defined in \req{x} and the function $f_\beta(\theta)$ defined by 
\bea
f_\beta(\theta)\equiv \sqrt{1+\sinh^2\beta \cos^2\theta}\, \quad {\rm with }\quad x f_\beta(\theta)<1 \,,
\eea
which follows from \req{beta}.
Making these replacements, the EMI takes the form
\begin{equation}
I_{\rm \ssc EMI}= \frac{8 \kappa_{(d)} 2^{2-d}x^{d-2}\pi^{d-2}}{\Gamma\!\(\frac{d-2}{2}\)^2}\cosh\beta \int_0^\pi \! \frac{\diff \theta_B\, \sin^{d-3}\theta_B }{ f_\beta\(\theta_B\)} \! \int_0^\pi \frac{ \diff \theta_A\, \, \sin^{d-3}\theta_A\cos\theta_A}{\(1-xf_{\beta}\(\theta_B\)\cos\theta_A\)^{d-2}}\, .
\end{equation} 
The integral over $\theta_A$ can be carried out explicitly, and a convenient expression for it reads
\bea
\int_0^\pi \frac{ \diff \theta_A\, \, \sin^{d-3}\theta_A \cos\theta_A}{\(1-xf_{\beta}\(\theta_B\)\cos\theta_A\)^{d-2}}&=&\frac{\sqrt{\pi} \, \Gamma\!\(\frac{d-2}{2}\)} {\Gamma\!\(\frac{d-1}{2}\)}\(\frac{d-2}{d-1}\) \frac{ x f_\beta\(\theta_B\)}{\[1-x^2f^2_\beta\(\theta_B\) \]^{d/2}} \nonumber \\ &&\qquad \qquad \qquad \times \,_2F_{1}\[\frac{d}{2},1,\frac{d+1}2;-\frac{x^2f^2_\beta\(\theta_B\)}{1-x^2f^2_\beta\(\theta_B\)}\] \,.
\eea
The final result is thus given as an integral over $\theta_B$, which we find convenient to express in terms of the variable $\xi\equiv \cos\theta_B$. Then, the final expression for the mutual information of two boosted spheres in the EMI model reads\footnote{To arrive at this expression we also use the identity
\bea
\,_2F_1\(a,b,c;z\)=(1-z)^{-a}\,_2F_1\[a,c-b,c;\frac{z}{z-1}\] \, .
\eea}
 \begin{equation}\label{EMI-final-1}
I_{\rm \ssc EMI}=2\gamma_d x^{d-1} \cosh\beta \int_0^1  \diff \xi \(1-\xi^2\)^{\frac{d-4}2}  \,_2F_{1}\[\frac{d}{2},\frac{d-1}2,\frac{d+1}2;x^2\(1+\sinh^2\beta \xi^2\)\]\, ,
\end{equation}
where 
\bea
\gamma_d\equiv \frac{2^4 \kappa_{(d)} \pi^{d-\frac{3}{2}}}{2^{d-1}\Gamma\(\frac{d-2}{2}\)  \Gamma\(\frac{d-1}{2}\) }\(\frac{d-2}{d-1}\)\,.
\eea
The result is a function of the cross ratio defined in \req{x} above and of the boost parameter $\beta$ introduced in \req{boost}.

For comparison purposes, it is important to write this back in terms of the $u$ and $v$ parameters. First from (\ref{chi1chi2-xbeta}) one can solve for $x$ and $\cosh\beta$ as a function of $\chi_1$ and $\chi_2$, 
\bea\label{xbetachi1chi2}
\frac{1}{x}=\frac{1}{\chi_2}+\frac{1}{\chi_1} \quad {\rm and} \quad \cosh\beta=\frac{1}{\chi_2}-\frac{1}{\chi_1}\, .
\eea
Using (\ref{uv}) we can rewrite these in terms of $u$ and $v$ as 
\bea\label{xbetauv}
x=\frac{\sqrt{u}}{1+\sqrt{v}}, \quad \cosh \beta=\frac{\sqrt{v}-1}{\sqrt{u}}\, .%\qquad 
\eea
%and 
%%\bea
%\alpha^2=\frac{(\sqrt{v}-1)^2-u}{u}\,.
%\eea
Notice that the trivial inequality $\cosh\beta>1$ implies 
$
\sqrt{v}\geq 1+\sqrt{u}
$, 
and the constraints $0\leq x\leq 1$ and $0\leq x\cosh\beta\leq 1$ are trivial consequences of this relation. %in these variables. 
In terms of $u$ and $v$ the final expression for the EMI reads 
 \bea\label{EMI-final-2}
&&I_{\rm \ssc EMI}=-2\gamma_d \(\frac{\sqrt{u}}{1+\sqrt{v}}\)^{d-2} \(\frac{1-\sqrt{v}}{1+\sqrt{v}}\)\nonumber \\
&&\qquad \qquad \times \int_0^1  \diff \xi \(1-\xi^2\)^{\frac{d-4}2}  \,_2F_{1}\[\frac{d}{2},\frac{d-1}2,\frac{d+1}2;\frac{\(1-\sqrt{v}\)^2\xi^2+u\(1-\xi^2\)}{\(1+\sqrt{v}\)^2}
\]\, .
\eea
%where 
%\bea
%c_d=\frac{2^4 \kappa_{(d)} \pi^{d-\frac{3}{2}}}{2^{d-1}\Gamma\(\frac{d-2}{2}\)  \Gamma\(\frac{d-1}{2}\) }\(\frac{d-2}{d-1}\)\,.
%\eea
Before diving into a detailed analysis of this result, let us study certain limiting cases.

\subsubsection{Long distance}\label{longdists}
Let us consider the long-distance behavior of \req{EMI-final-1} ---or equivalently, \req{EMI-final-2}. To this end we need to go back to the disjoint-spheres geometry. In particular, the relation between the parameters $x$ and $\cosh\beta$, and $\chi_1$ and $\chi_2$, which was derived assuming the causal cones of the spheres to be nested, needs to be modified. The modification entails a simple interchange between $\chi_1$ and $\chi_2$. Indeed, for disjoint spheres, (\ref{xbetachi1chi2}) becomes
\bea\label{xbeta}
\frac{1}{x}=\frac{1}{\chi_2}+\frac{1}{\chi_1} \quad {\rm and} \quad \cosh\beta=\frac{1}{\chi_1}-\frac{1}{\chi_2}\,, 
\eea
and (\ref{xbetauv}) is now 
\bea\label{newxbetauv}
x=\frac{\sqrt{u}}{1+\sqrt{v}}, \quad \cosh \beta=\frac{1-\sqrt{v}}{\sqrt{u}}\,,\qquad \text{with} \qquad \sqrt{u}\leq 1-\sqrt{v}\,.
\eea
%with 
%\bea
%\sqrt{u}\leq 1-\sqrt{v}\,.
%\eea
To derive the leading behavior of the EMI in such a limit, we would need to study $x$ and $\cosh\beta$ for $L\gg R_A, R_B$ using eqs. (\ref{xbeta}) and (\ref{chi1chi2-disconc}).  The relevant leading expressions are thus (see for instance section (2.2) of \cite{casini2021mutual})
\bea\label{xbetaleading}
x\sim \frac{2R_A R_B}{L^2}\,, \quad {\rm and }\quad \cosh\beta\sim \[2(n_A\cdot l)( n_B\cdot l)-n_A\cdot n_B\]\,.
\eea

In the long distance limit $x\ll 1$ and $\cosh \beta\sim \mathcal{O}(1)$. As a consequence, the leading behavior of \req{EMI-final-1} in this regime can be obtained by setting $x=0$ in the integrand, which makes the hypergeometric function equal to one, and replacing everywhere else the relations (\ref{xbetaleading}). This leads to 
\bea
I_{\rm \ssc EMI} \overset{(L\gg R_A,R_B)}{=}\frac{2^5 \kappa_{(d)} \pi^{d-\frac{3}{2}}\(\frac{d-2}{d-1}\)}{2^{d-1}\Gamma\(\frac{d-2}{2}\)  \Gamma\(\frac{d-1}{2}\) }\( \frac{2R_A R_B}{L^2}\)^{d-1} \[2(n_A\cdot l)( n_B\cdot l)-n_A\cdot n_B\]\int_0^1  \diff  \xi \(1-\xi^2\)^{\frac{d-4}2} \, .
\eea
The remaining integral can be trivially done, 
 \bea\label{int-zero-boost}
\int_0^1  \diff \xi \(1-\xi^2\)^{\frac{d-4}2}=\frac{\sqrt{\pi}}{2}\frac{\Gamma\(\frac{d-2}{2}\)}{\Gamma\(\frac{d-1}{2}\)}\,,
\eea
leading to the final result presented in \req{eee} which, as mentioned above,
% The final result is thus
 %\bea \label{Long-dist-EMI}
%I_{\rm \ssc EMI}(A,B)\overset{}{=}\frac{4 \pi^{d-1}(d-1)(d-2)}{\(\Gamma\(\frac{d+1}{2}\)\)^2} \kappa_{(d)} \, \(2(n_A\cdot l)( n_B\cdot l)-n_A\cdot n_B\) \(\frac{R_A R_B}{L^2}\)^{d-1}
 %\eea
 agrees ---up to an undetermined overall constant--- with the long-distance behavior of the mutual information dominated by a free fermion.

We can similarly study (\ref{EMI-final-2}) (with an overall minus sign change due to the new relations (\ref{xbetauv})). In that case, the analogous leading expressions for $u$ and $v$ are  
\bea
u\sim \frac{16 R_A^2 R_B^2}{L^4}+\mathcal{O}(L^{-5}) \,, \quad {\rm and }\quad v\sim 1-\frac{8 R_A R_B}{L^2}\[2(n_A\cdot l)( n_B\cdot l)-n_A\cdot n_B\]+\mathcal{O}(L^{-3}) \,.
\eea
and plugging these into \req{EMI-final-2} will lead to \req{eee}.

Since $u$ and $v$ are in general independent variables, we can study other limits. For instance, we can take $u\to 0$ while leaving $v$ untouched, or take $v\to 1$ with $u$ arbitrary, among others. Let us mention the $u\to 0$ case with arbitrary $v$, as it turns out to have a simple analytic form.
%\begin{itemize}
%\item $u\to 0$, $v$ arbitrary. 
Indeed, in this limit the integral in (\ref{EMI-final-2}) can be carried out explicitly, leading to 
\bea
 &&\int_0^1  \diff \xi \(1-\xi^2\)^{\frac{d-4}2}  \,_2F_{1}\[\frac{d}{2},\frac{d-1}2,\frac{d+1}2;\(\frac{1-\sqrt{v}}{1+\sqrt{v}}\)^2\xi^2\]\nonumber \\
&& \qquad \qquad \qquad \qquad \qquad \qquad  =\frac{4(d-1)\sqrt{\pi}\Gamma\(\frac{d-2}{2}\)}{\Gamma\(\frac{d+1}{2}\)} \,_2F_{1}\[\frac{1}{2},\frac{d}2,\frac{d+1}2;\(\frac{1-\sqrt{v}}{1+\sqrt{v}}\)^2\]\, ,
\eea
and thus the EMI becomes
 \bea\label{long-d-uv}
I_{\rm \ssc EMI}\overset{u\rightarrow 0}{=} \frac{4 \pi^{d-1}(d-1)(d-2)}{2^{d-1}\(\Gamma\(\frac{d+1}{2}\)\)^2} \kappa_{(d)} \(\frac{\sqrt{u}}{1+\sqrt{v}}\)^{d-2} \(\frac{1-\sqrt{v}}{1+\sqrt{v}}\) \,_2F_{1}\[\frac{1}{2},\frac{d}2,\frac{d+1}2;\(\frac{1-\sqrt{v}}{1+\sqrt{v}}\)^2\]\,.
\eea
%\end{itemize}

Expression (\ref{long-d-uv}) can be compared directly with the leading contributing conformal block. For $u\to 0$ and $v$ arbitrary, this reads \cite{Dolan:2011dv}
\bea
G^d_{d-1,1}(u,v)\overset{u\rightarrow 0}{=} u^{\frac{d-2}{2}}(1-v)\,_2F_{1}\[\frac{d}{2},\frac{d}2,d;1-v\] \, ,
\eea
whose functional dependence coincides exactly with (\ref{long-d-uv}) after the use of some identities.\footnote{ We found the following identity numerically,
\bea
 4\(\frac{2}{1+\sqrt{v}}\)^{d-2} \(\frac{1-\sqrt{v}}{1+\sqrt{v}}\) \,_2F_{1}\[\frac{1}{2},\frac{d}2,\frac{d+1}2;\(\frac{1-\sqrt{v}}{1+\sqrt{v}}\)^2\]=(1-v)\,_2F_{1}\[\frac{d}{2},\frac{d}2,d;1-v\]\, .
\eea
Perhaps this follows from standard hypergeometric identities. Otherwise, it would be interesting to prove it rigorously. 
} Thus, comparing with the expected behavior of  $I_{\rm \ssc EMI}(A,B) $ at long distances, this is
\bea
I_{\rm \ssc EMI}(A,B)|_{u\rightarrow 0}= b^{\ssc \rm (ferm)}_{d-1,1}G^d_{d-1,1}(u,v)|_{u\rightarrow 0}\, ,
\eea
where $b_{d-1,1}^{\ssc \rm (ferm)}$ was defined in \req{bf}, we can read off the required value for $\kappa_{(d)}$. The result appears in \req{kappad} above.
%\bea\label{kappa-d}
%\kappa_{(d)}&=& \frac{4^{d-1}\(\Gamma\(\frac{d+1}{2}\)\)^2} {4 \pi^{d-1}(d-1)(d-2)}b^{(f)}_{d-1,1}\,, \\
%\kappa_{(d)}&=&\frac{2^{\[\frac{d}2\]}\Gamma\(d-2\)\(\Gamma\(\frac{d+1}{2}\)\)^2}{8\pi^{d-\frac32}\Gamma\(d+\frac{1}{2}\)}\,,
%\eea
%where we used the value of $b^{(f)}_{d-1,1}$ from (\ref{bf}).
Already at this point it is puzzling
%However, at this point it should be puzzling 
that the $u\to 0$ limit of (\ref{EMI-final-2})  agrees exactly with the contribution of the leading conformal block away from $v\to 1$. In other words, it is surprising to find a perfect agreement away from the usual long-distance limit.  %although this limit is related to it. 

\subsubsection{Zero relative boost}
An important limiting case corresponds to $\cosh\beta=1$, which describes concentric spheres lying on the same space-like hyperplane ---which we can take to be the $t=$constant surface. In that case, we can write down a closed analytic expression for $I_{\rm \ssc EMI}(A,B)$ for all $d$.  First, in that limit, the argument of the hypergeometric function in (\ref{EMI-final-1}) becomes independent of the integration variable and one finds
 \begin{equation}\label{EMI-zero-boost}
I_{\rm \ssc EMI}(A,B)=\frac{4 \pi^{d-1}(d-1)(d-2)}{2^{d-1} \Gamma\[\frac{d+1}{2}\]^2} \kappa_{(d)}\, x^{d-1} \,_2F_{1}\[\frac{d}{2},\frac{d-1}2,\frac{d+1}2;x^2\]\,,
\end{equation}
where $x$ can be determined in terms of the geometric parameters for disjoint spheres on the same hyperplane, using (\ref{xbeta}) and (\ref{chi1chi2-disconc}). For spheres on the same hyperplane it is convenient to define the vector that joins the spheres' centers, 
\bea
D\equiv Ll +R_Bn_B-R_An_A\,,
\eea
since now, $n_A\cdot D=0$, $n_B\cdot D=0$ and $n_A\cdot n_B=-1$. Trading $L$ for $D$, $\chi_1$ and $\chi_2$ take the form
\bea
\chi_1=\frac{4R_AR_B}{|D|^2-\(R_A-R_B\)^2}\, , \quad \chi_2=\frac{4R_AR_B}{|D|^2-\(R_A+R_B\)^2}\, ,
\eea
one can check using (\ref{xbeta}) that $\cosh\beta=1$ as we wanted, and 
\bea
x=\frac{2R_AR_B}{|D|^2-R_A^2-R_B^2}\,.
\eea
Plugging this expression into (\ref{EMI-zero-boost}) gives us the EMI mutual information as a function of the physical parameters. 

As described at the beginning of the section, we would like to subtract from this expression the contribution of the conformal block associated to the spin $1$ conserved current made out of free fermions, namely, the term
$
b^{\ssc \rm (ferm)}_{d-1,1}G^d_{d-1}(u,v)
$. 
The limit considered above, $\cosh\beta=1$, corresponds in the $u$ and $v$ variables to the case in which 
\bea\label{urelv}
\sqrt{u}=1-\sqrt{v}\, ,
\eea
which follows from (\ref{newxbetauv}). However, it turns out that in the CFT literature these conformal blocks are often written in terms of yet another set of variables (parameters), $\{y, z\}$ which are defined via the relations
\bea\label{yz}
u= yz, \quad {\rm and}\quad v=(1-y)(1-z)\,.
\eea
In terms of these variables, the constraint equation (\ref{urelv}) implies 
\bea
y=z\,.
\eea 
This is called the ``diagonal limit'' of the conformal blocks \cite{Hogervorst:2013kva}. In such limit, $u$ and $v$ become simple functions of $z$
\bea
u(z)=z^2\,,\qquad {\rm and } \qquad v(z)=(1-z)^2\,.
\eea

Now, a close formula for a conformal block of arbitrary scaling dimension and spin was derived in \cite{ElShowk:2012ht}. For $\Delta=d-1$ and $J=1$ this reads
\bea\label{Conf-block-J=1}
G^d_{d-1,1}(u(z),v(z))\equiv G^d_{d-1,1}(z)=\frac{2-z}{2z} \left(\frac{z^2}{1-z} \right)^{\frac{d}{2}}{}_{2}F_1 \left[1,\frac{d}{2},\frac{d+1}{2};\frac{z^2}{4(z-1)} \right]\,.
\eea
Hence, our goal is then to subtract from (\ref{EMI-zero-boost}) the function %$b^{\rm \ssc (ferm)}_{d-1,1}G^d_{d-1,1}(z)$ 
\bea
b^{\ssc \rm (ferm)}_{d-1,1}G^d_{d-1,1}(z)=2^{[\frac{d}2]+1}\frac{\sqrt{\pi}\, \Gamma\(d\)}{4^d\, \Gamma\(d+\frac{1}{2}\)}\frac{2-z}{2z} \left(\frac{z^2}{1-z} \right)^{\frac{d}{2}} {}_{2}F_1 \left[1,\frac{d}{2},\frac{d+1}{2};\frac{z^2}{4(z-1)} \right]\, ,
\eea
for which we need to rewrite \req{EMI-zero-boost} as a function of $z$.
%For that purpose we need to rewrite (\ref{EMI-zero-boost}) as a function of $z$. 
The relation between $x$ and $z$ can be easily derived from (\ref{newxbetauv}) and yields
\bea
x=\frac{\sqrt{u(z)}}{1+\sqrt{v(z)}}=\frac{z}{2-z}\, .
\eea
Plugging this into (\ref{newxbetauv}) and using the calibrated value of the parameter $\kappa_{(d)}$ (\ref{kappad}), we obtain 
\bea
I_{\rm \ssc EMI}(A,B)=2^{[\frac{d}2]+1}\frac{\sqrt{\pi}\, \Gamma\(d\)}{4^d\, \Gamma\(d+\frac{1}{2}\)}\frac{2-z}{2z} \left(\frac{z^2}{1-z} \right)^{\frac{d}{2}} {}_{2}F_1 \left[1,\frac{d}{2},\frac{d+1}{2};\frac{z^2}{4(z-1)} \right]\,,
\eea
which is identical to $b^{\ssc \rm (ferm)}_{d-1,1}G^d_{d-1,1}(z)$.
This is a rather surprising result. For spherical regions, the EMI exactly equals the conformal block contribution of a spin $1$ conserved current made out of a free fermion field. In other words, no other operators contribute to the sum in the RHS of (\ref{I-emi}), and the iterative process mentioned at the beginning of the section is completed after the very first step.

\subsubsection{General case for particular dimensions\label{313}}

At this point it is obvious that in the most general case we must have the equality 
\bea
I_{\rm \ssc EMI}(A,B)=b^{\rm \ssc (ferm)}_{d-1,1}G^d_{d-1,1}(u,v)\,,
\eea
and indeed, there is an elegant way to prove this equivalence using a smeared representation of the so-called OPE blocks. The reader interested in seeing such derivation can find it in Appendix \ref{B}.

It is instructive to write down closed-form expressions for each conformal block by explicitly computing the integral in (\ref{EMI-final-1}) for the various dimensions. We rewrite the formula here for convenience
 \begin{align}\notag
I_{\rm \ssc EMI}(A,B)&=\frac{2^d\, \Gamma\[\frac{d-1}2\] b^{\rm \ssc (ferm)}_{d-1,1}}{\sqrt{\pi}\Gamma\[\frac{d-2}{2}\] } x^{d-1} \cosh\beta \\ &  \times\int_0^1  \diff \xi \(1-\xi^2\)^{\frac{d-4}2}  \,_2F_{1}\[\frac{d}{2},\frac{d-1}2,\frac{d+1}2;x^2\(1+\sinh^2\beta \xi^2\)\]\, ,\label{EMI-integral-3}
\end{align}
where we have included explicitly the factor $b^{\rm \ssc (ferm)}_{d-1,1}$ to facilitate comparison with the conformal blocks. Its dependence on the physical parameters $R_A$, $R_B$, $L$ and the unit vectors $n_A$, $n_B$ and $l$ can be obtained from \req{xbeta} and the definitions (\ref{chi1chi2-disconc}). Let us study now the above expression for $d=3,4,5,6$.
 %Now, we will study the mutual information for the following particular dimensions:
%\begin{itemize}
%\item {\bf $d=3$}
%\end{itemize}

{\bf Three dimensions.} 
 In this case, the integral in \req{EMI-integral-3} gives
 \bea\label{Elliptic-int}
 \int_0^1  \frac{\diff \xi  }{\sqrt{1-\xi^2}}\,_2F_{1}\[1,\frac{3}{2},2;x^2\(1+\sinh^2\beta \xi^2\)\]  =\frac{2}{x^2}\[\frac{\Pi\(-\sinh^2\beta \Big|\frac{\sinh^2\beta x^2}{1-x^2}\)}{\sqrt{1-x^2}}-\frac{\pi}{2\cosh \beta}\]\, ,
 \eea
 where $\Pi\(n|m\)$ is the complete elliptic integral of third kind\footnote{We adopt the following conventions for the elliptic integrals used in this paper. The elliptic integral of first kind K$(n)$, second kind E$(m)$ and third kind $\Pi\(n|m\)$ are defined as:
 \bea\label{KE}
{\rm K}\(n\)=\int_0^{\frac\pi2}\frac{\diff \theta}{\sqrt{1-n\sin^2\theta}}\, ,\qquad \qquad  {\rm E}\(m\)=\int_0^{\frac\pi2}\sqrt{1-m\sin^2\theta}\,\diff \theta\, ,
 \eea
 and
 \bea\label{Pi}
\Pi\(n|m\):=\int_0^{\frac\pi2}\frac{\diff \theta}{\(1-n\sin^2\theta\)\sqrt{1-m\sin^2\theta}}\,.
\eea
}
and its second argument in (\ref{Elliptic-int}) is bounded as
\bea
0\leq \frac{\sinh^2\beta x^2}{1-x^2} \leq 1\,.
\eea
%This combination will appear in several places below, so we give it a name,
%Since this same combination will appear later on, we will give it a name:
%\bea
%\chi^2\equiv \frac{\sinh^2\beta x^2}{1-x^2}\,.
%\eea
Hence, the result for the mutual information is
\bea\label{EMI-d=3}
I_{\rm \ssc EMI}(A,B)=\frac{16}{\pi}\, b^{\ssc \rm (ferm)}_{2,1} \[\frac{\cosh\beta}{\sqrt{1-x^2}}\, \Pi\(-\sinh^2\beta \Big|\frac{\sinh^2\beta x^2}{1-x^2}\)-\frac{\pi}{2}\]\, .
\eea
To compare with the CFT literature it is convenient to write the expression in terms of the $\{y,z\}$ variables defined in (\ref{yz}). Here we write each term appearing in (\ref{EMI-d=3}) instead. We have\footnote{For $0\leq y,z\leq 1$ the expressions can be written as 
\bea
\sinh^2\beta=\frac{\(\sqrt{1-y}-\sqrt{1-z }\)^2}{yz }\,,\qquad \frac{\sinh^2\beta x^2}{1-x^2}=\frac{\(\sqrt{1-y}-\sqrt{1-z }\)^4}{\(y-z \)^2}\,.
\eea
However, in some applications it is important not to make such assumption since one would need the expression to be valid in the whole complex plane. }
\bea \label{someyz}
&&\sinh^2\beta=\frac{2-y-z-2\sqrt{(1-y)(1-z )}}{yz }\,,\qquad \frac{\sinh^2\beta x^2}{1-x^2}=\frac{\(2-y-z-2\sqrt{(1-y)(1-z )}\)^2}{\(y-z \)^2}\,, \nonumber \\
&& \qquad \frac{\cosh\beta}{\sqrt{1-x^2}}=\frac{yz -y-z }{|y-z |} \sinh\beta \,.
\eea
In $d=3$ there is no general closed-form expression for the corresponding conformal block to compare with, although the interested reader can find an integral representation in \cite{Dolan:2011dv}, which can be shown to be equivalent to (\ref{someyz}) for $\Delta=2$ and $J=1$. For comparison purposes, however, we prefer to turn the series representation of  \cite{Hogervorst:2013sma} for $d=3$ into an integral representation ---see \req{current-CB} below--- and numerically compare the results of both expressions. We find a perfect match.
 
%\begin{itemize}
%\item {\bf $d=4$}
%\end{itemize}

{\bf Four dimensions.} In this case, the relevant integral can also be done explicitly and its answer turns out to be given in terms of elementary functions, namely, 
\bea
 \int_0^1  \diff \xi  \,_2F_{1}\[2,\frac{3}2,\frac{5}2;x^2\(1+\sinh^2\beta \xi^2\)\] =\frac{3}{2 x^3}\[\frac{\tanh^{-1}\left[\frac{ \sinh\beta x}{\sqrt{1-x^2}}\right]}{\sinh\beta\, \sqrt{1-x^2} }-\frac{\tanh^{-1}\left[ x \cosh\beta \right]}{\cosh\beta}\]\, .
\eea
Therefore, the mutual information is 
\bea
I_{\rm \ssc EMI}(A,B)&=&12\,b^{\rm \ssc (ferm)}_{3,1}\[\frac{\cosh \beta}{\sinh\beta\, \sqrt{1-x^2}}\tanh^{-1}\left[\sqrt{\frac{\sinh^2\beta x^2}{1-x^2}} \right]- \tanh^{-1}\left[ x \cosh\beta \right]\]\,.
\eea
This result can be expressed more simply in terms of the $\{y,z\}$ variables. We write the various pieces explicitly,
\bea\label{rels-xa-yz}
&&\tanh^{-1}\left[ x \sqrt{1+\cosh^2\beta} \right]=\frac{1}{4}\log\[(1-y)(1-z)\],\nonumber \\
&& \tanh^{-1}\left[\sqrt{\frac{\sinh^2\beta x^2}{1-x^2}} \right]
%=\frac{1}{2}\log\(\frac{1+\chi}{1-\chi}\)
=\frac12\log\(\frac{2-y-z+|y-z|}{2\sqrt{\(1-y\)\(1-z\)}}\)\,.
\eea
Using \req{someyz}, and after some simplifications we get for the mutual information:
\bea\label{EMI-d=4}
I_{\rm \ssc EMI}(A,B)&=&3 b^{\ssc \rm (ferm)}_{3,1} \[\frac{y z-y-z}{(y-z)}\log\(\frac{1-z}{1-y}\)-\log\[(1-y)(1-z)\]\]\,.
\eea
Indeed, for $d=4$, earlier work on conformal blocks provided a general formula for the conformal block associated to any operator with arbitrary scaling dimension and spin. Evaluating $\Delta=d-1$ and $J=1$ in the general formula ---see \eg  eq. (7.1) of \cite{Dolan:2000ut}--- one gets 
\bea
G^4_{3,1}(y,z)=3\[\frac{y z-y-z}{(y-z)}\log\(\frac{1-z}{1-y}\)-\log\[(1-y)(1-z)\]\]\,,
\eea
in perfect agreement with our results.

%\begin{itemize}
%\item {\bf $d=5$}
%\end{itemize}

%\comment{HEREEEEE}

{\bf Five dimensions.} In this case one needs to work a bit harder to find a closed-form expression. However, the final answer can also be written compactly in terms of Elliptic integrals of the first, second and third kinds. %and it is nicely expressed in terms of $x$ and $\cosh\beta$ (remember $\alpha=\sinh\beta$).
 The relevant integral is
 \bea\label{Elliptic-int-2} 
&&\int_0^1 \diff \xi \,\sqrt{1-\xi^2}  \,_2F_{1}\[2,\frac{5}{2},3;x^2\(1+\sinh^2\beta\, \xi^2\)\] \nonumber \\ 
&&\qquad \qquad =\frac{2}{3x^4 \cosh^4\beta}\Bigg[\pi \cosh\beta
 +\frac{2}{\sqrt{1-x^2\cosh^2\beta}} 
 \Bigg\{ \frac{\cosh^2\beta}{\sinh^2\beta}\Bigg[{\rm E}\(\frac{x^2 \sinh^2\beta}{x^2 \cosh^2\beta-1}\)  \nonumber \\
 && \qquad \qquad \qquad \qquad  -{\rm K}\(\frac{x^2 \sinh^2\beta}{x^2 \cosh^2\beta-1}\)\Bigg]
% &&\qquad \qquad \qquad \qquad \qquad \qquad \qquad \qquad \qquad \qquad \qquad 
-\Pi\(\frac{\sinh^2\beta}{\cosh^2\beta} \Bigg|\frac{x^2 \sinh^2\beta}{x^2 \cosh^2\beta-1} \) \Bigg\}\Bigg]\,,
 \eea
 where K, E and $\Pi$ are the elliptic integrals of first, second and third kind ---the conventions are described in eqs. (\ref{KE}) and (\ref{Pi}).
Since the above is a rather complicated formula, we checked that it reduces to the right expression for $\beta\to0$. Our formula for the EMI in $d=5$ is therefore
\bea\label{EMI-d=5}
&&I_{\rm \ssc EMI}(A,B)=\frac{128}{3\pi}\,\frac{ b^{\rm \ssc(ferm)}_{4,1}}{\cosh^3\beta} 
\Bigg[\pi \cosh\beta+\frac{2}{\sqrt{1-x^2\cosh^2\beta}} \\
 &&
 \times \Bigg\{ \frac{\cosh^2\beta}{\sinh^2\beta}\[{\rm E}\(\frac{x^2 \sinh^2\beta}{x^2 \cosh^2\beta-1}\)-{\rm K}\(\frac{x^2 \sinh^2\beta}{x^2 \cosh^2\beta-1}\)\] -\Pi\(\frac{\sinh^2\beta}{\cosh^2\beta} \Bigg|\frac{x^2 \sinh^2\beta}{x^2 \cosh^2\beta-1} \)\Bigg\}\Bigg]\, , \nonumber
 \eea
where the arguments of the Elliptic functions in terms of the $\{y,z\}$ variables have the form
\bea
&&\frac{\sinh^2\beta}{\cosh^2\beta}=\frac{2-y-z-2\sqrt{(1-y)(1-z )}}{2-y-z-2\sqrt{(1-y)(1-z )}+yz}\,, \nonumber\\
&&\frac{x^2 \sinh^2\beta}{x^2 \cosh^2\beta-1}=-\frac{2-y-z-2\sqrt{(1-y)(1-z )}}{4\sqrt{(1-y)(1-z )}}\,.
\eea
We can rewrite the other terms using the relations
\bea
x=\frac{\sqrt{y \,z}}{1+\sqrt{(1-y)(1-z)}}\,, \qquad{\rm and }\qquad \sinh^2\beta =\frac{2-y-z-2\sqrt{(1-y)(1-z )}}{yz}\, ,
\eea
respectively. Similarly to the $d=3$ case, due to the lack of a closed-form expression for the corresponding conformal block, we numerically compare our result with the ones of 
%As done for the $d=3$ case, here again, due to the lack of a closed-form expression as the one we present here, we numerically compare our result with the results of 
appendix \ref{CB-CC}, finding perfect agreement. 

%\begin{itemize}
%\item {\bf $d=6$}
%\end{itemize}
{\bf Six dimensions}. Once again, the resulting integral can be carried out explicitly, leading to
\bea\label{int-d=6}
&&\int_0^1  \diff \xi \(1-\xi^2\) \,_2F_{1}\[\frac{5}2,3,\frac{7}2;x^2\(1+\sinh^2\beta\xi^2\)\] \nonumber \\  
&&\qquad \qquad  = \frac{5}{8 x^5}\Bigg[
 \frac{1-2\sinh^2\beta+x^2\(1-3\sinh^2\beta\)}{\sinh^3\beta\, \(1-x^2\)^{\frac32} }\tanh^{-1}\left[\frac{ \sinh\beta x}{\sqrt{1-x^2}}\right] \nonumber  \\
  &&\qquad \qquad   \qquad \qquad  \qquad   \qquad \qquad  -2\frac{\tanh^{-1}\left[ x \cosh\beta \right]}{\cosh\beta}-\frac{x}{(1-x^2)\sinh^2\beta}\Bigg]\,.
\eea
Using relations (\ref{rels-xa-yz}), plus
\bea 
\cosh\beta \[\frac{1-2\sinh^2\beta+x^2\(1-3\sinh^2\beta\)}{\sinh^3\beta\, \(1-x^2\)^{\frac32} }\]=-\frac{2(y+z-yz)\(y^2(1+z)+z^2(1+y)-4yz\)}{|y-z|^3}\, ,
\eea
and 
\bea
\frac{\cosh\beta\, x}{(1-x^2)\sinh^2\beta}=\frac{y z (y+z-yz)}{(y-z)^2}\,,
\eea
we can write down the expression for the mutual information in terms of the $\{y,z\}$ variables explicitly, 
\bea\label{EMI-d=6}
I_{\rm \ssc EMI}(A,B)&=&15\, b^{\rm \ssc (ferm)}_{5,1} \Bigg[\frac{\(y z-y-z\)\(y^2(1+z)+z^2(1+y)-4yz\)}{(y-z)^3}\log\(\frac{1-z}{1-y}\) \nonumber \\
&&\qquad \qquad \quad +\frac{2y z (yz-y-z)}{(y-z)^2}-\log\[(1-y)(1-z)\]
\Bigg]\,.
\eea
The result for the general conformal block for $d=6$ was originally derived in \cite{Dolan:2011dv} ---\eg see their eq. (5.14). In the case of interest here,  their result for $G^6_{5,1}(y,z)$ reduces to $I_{\rm \ssc EMI}(A,B)/b^{\rm \ssc (ferm)}_{5,1} $, in perfect agreement with our general result.
%\bea
%G^6_{5,1}(y,z)&=&15\,\Bigg[\frac{\(y z-y-z\)\(y^2(1+z)+z^2(1+y)-4yz\)}{(y-z)^3}\log\(\frac{1-z}{1-y}\) \nonumber \\
%&&\qquad \qquad+\frac{2y z (yz-y-z)}{(y-z)^2}-\log\[(1-y)(1-z)\]
%\Bigg]\,.
%\eea
%in perfect agreement with our result.

\section{Free fermion at long distances for arbitrary regions}
\label{longdist}

As mentioned earlier, for the EMI model the mutual information of two entangling regions $A$, $B$ is given, in the long-distance limit, by the expression
\begin{equation}\label{emilong}
I_{\rm \ssc EMI}(A,B)\overset{}{=} 4(d-1)(d-2)\kappa_{(d)}\frac{\text{vol}(A)\cdot \text{vol}(B)}{r^{2(d-1)}}+  \mathcal{O} (r^{-{2d+1}}) \, .
\end{equation}
Here, $\kappa_{(d)}$ is the usual constant characteristic of the model, which is therefore completely independent of the geometry of $A$ and $B$.  The expression analogous to \req{emilong} for a free fermion reads \cite{Casini:2009sr}
%At long distances, the mutual information of two regions $A$,$B$ for a free fermion field behaves as
\begin{equation}
I_{\rm \ssc ferm}(A,B) = g(A,B) \cdot \frac{\text{vol}(A)\cdot \text{vol}(B)}{r^{2(d-1)}}+  \mathcal{O} (r^{-{2d+1}}) \, ,
\end{equation}
where 
%where ``vol$(A)$'' here stands for the area of the entangling region $A$ and 
$g(A,B)$ is a number which, in principle, may depend on the geometry of the regions. Given the match in the scaling, it is natural to wonder whether the free fermion actually coincides exactly with the EMI in this regime. Where in this case, the function $g(A,B)$ would in fact be given by a fixed constant, independent of the geometric details of $A$ and $B$.

 In order to test this possibility, we will compute here the mutual information for different shapes for a free fermion. From \req{fer} we can analytically extract  $g(A,B)$ in the case of two disk regions for the free fermion. We have
 \begin{equation}\label{gdisksf}
 g(A,B)|_{\rm \ssc spheres} =\frac{2^{[\frac d2]-1} \, \Gamma\!\(d\) \,\(\Gamma\!\(\frac{d+1}2\)\)^2}{\pi^{d-\frac32}\,\Gamma\!\(d+\frac12\)}\, .
 \end{equation}

A useful expression for  $g(A,B)$ was obtained in \cite{Casini:2009sr} in terms of the resolvent of the vacuum correlator of the Dirac field restricted to each region, $A$ and $B$. For the two-point function $C(x,y)\equiv \braket{\Psi(x)\Psi(y)^{\dagger}}$, 
% The correlator is given by
%\begin{equation}
%C(x,y)\equiv \braket{\Psi(x)\Psi(y)^{\dagger}}=\frac{1}{2}\delta(x-y)+i c_3  \gamma^i \gamma^0 \frac{(x-y)^i}{|x-y|^3}\, ,
%\end{equation}  
%where $c_3=2\pi$ \comment{check coefficients}
%and 
the resolvent is defined as $R(\beta)\equiv [C+\beta -1/2]^{-1}$. Then, for regions which are mirror symmetric of themselves with respect to the line which separates $A$ and $B$, we have %\comment{more on where this formula comes from?}
\begin{equation}\label{gbbs}
g(A,B)\text{vol}(A)\cdot \text{vol}(B)=\frac{\(\Gamma\!\(\frac d2\)\)^2}{4 \pi^d} \int_{1/2}^{\infty} \diff \beta (\beta-1/2) \left[ \tr \left[ \overline{ R_A}(\beta) \overline{ R_B^2}(\beta)+ \overline{ R_B}(\beta)\overline{R_A^2}(\beta) \right]- (\beta \leftrightarrow - \beta) \right] \, ,
\end{equation}
where $\overline{R_A}(\beta) \equiv \int_{x\in A}\int_{y\in A} R(\beta;x,y) $ denotes sum over the spatial variables belonging to region $A$.

Using this formula we compute the coefficient $g(A,B)$ in $d=2+1$ in the lattice, and test the possible invariance with the shape of the regions. From \req{gdisksf} we have 
\begin{equation}\label{gdisksf3}
 g(A,B)|^{d=3}_{\rm \ssc disks} =\frac{16}{15\pi^2}\, .
 \end{equation}
This will also allow us to test the precision of our numerical calculations.

The formula (\ref{gbbs}) can be exploited in the lattice as follows. First, let $C_{A,ij}^{\alpha \beta}\equiv \braket{\psi_i^{\alpha} {\psi_j^\dagger}^{\beta}}$ be the correlator for lattice fermionic fields corresponding to $i,j=1,\dots,N_A$ where $N_A$ is the number of lattice sites of $A$ and where we denoted spinorial indices by $\alpha,\beta=0,1$.
Then, if we denote by $u_{A,j}^{(\lambda_A) \beta}$ the eigenvector of $C_{A,ij}^{\alpha \beta}$ corresponding to a given eigenvalue $\lambda_A$, we can write
\begin{equation}
\sum_{j,\beta} C_{A,ij}^{\alpha\beta} u_{A,j}^{(\lambda_A) \beta}=\lambda_A\, u_{A,i}^{(\lambda_A) \alpha}\, , \quad C_{A,ij}^{\alpha\beta}= \sum_{\lambda_A} \lambda_A u_{A,i}^{(\lambda_A) \alpha} {u_{A,j}^{(\lambda_A) \beta}}^{\dagger}\, .
\end{equation}
Inserting the second expression in \req{gbbs} and using  
\begin{align}\notag
\int_{1/2}^{\infty} \diff \beta (\beta-1/2) \Big[& \frac{1}{(\lambda_A +\beta -1/2)(\lambda_B+\beta-1/2)^{2}} + \frac{1}{(\lambda_B +\beta -1/2)(\lambda_A+\beta-1/2)^{2}}  \\ &-(\beta \leftrightarrow - \beta)  \Big]= f(\lambda_A,\lambda_B)\, ,
\end{align}
where
\begin{equation}
f(\lambda_A,\lambda_B) \equiv \frac{1}{(\lambda_A - \lambda_B)} \log \left[ \frac{\lambda_A (1-\lambda_B)}{\lambda_B (1-\lambda_A)} \right]\, ,
\end{equation}
we find the formula
\begin{equation}\label{ggg}
g(A,B)\,\text{vol}(A)\cdot \text{vol}(B)=\frac{1}{16\pi^2} \sum_{\lambda_A,\lambda_B}  f(\lambda_A,\lambda_B)\, \left| {v_{B}^{(\lambda_B)}}^{\dagger} \cdot v_{A}^{(\lambda_A)} \right|^2 \, , \quad \text{where} \quad  v_{A}^{(\lambda_A)\, \alpha}\equiv \sum_i u_{A,i}^{(\lambda_A) \, \alpha}\, ,
\end{equation}
and, for fixed $\lambda_A$ and $\lambda_B$, ${v_{B}^{(\lambda_B)}}^{\dagger} \cdot v_{A}^{(\lambda_A)}$ denotes a scalar product between two-component vectors corresponding to the spinorial indices.
In words, given the correlators matrix corresponding to each region, we first obtain its eigenvalues and eigenvectors. Using the former we can evaluate the function $f(\lambda_A,\lambda_B)$ for every pair of eigenvectors. Then, for each spinorial index $\alpha=0,1$, we sum over the lattice sites of each region in order to obtain the $v^{(\lambda),\alpha}$. Finally, we use \req{ggg} to evaluate $g(A,B)$.\footnote{The analogous expression for the long-distance coefficient of the MI for scalar fields was obtained in \cite{shiba2020direct}. For fermions, the same formula (\ref{ggg}) applies in the lattice without correction for fermion doubling. The reason is that the eigenvectors are averaged over position and this eliminates the doubling modes which have large momentum $\sim \pi$ in the Brillouin zone.}

The three-dimensional lattice Hamiltonian for the free fermion reads
\begin{equation}
H=-\frac{i}{2} \sum_{n ,m} \left[  \left(\psi^{\dagger}_{m, n} \gamma^0   \gamma^1 (\psi_{m+1,n}-\psi_{m,n})+\psi^{\dagger}_{m,  n} \gamma^0 \gamma^2   (\psi_{m,n+1}-\psi_{ m,n}  ) \right)  - h.c.\right] \, ,
\end{equation}
and the vacuum-state correlators are given in this case by \cite{Casini:2009sr}
\begin{equation}
C_{(n,k),(j,l)} = \frac{1}{2}\delta_{n,j} \delta_{kl}- \int_{-\pi}^{\pi} \diff x  \int_{-\pi}^{\pi} \diff y \frac{\sin (x) \gamma^0 \gamma^1+\sin(y) \gamma^0 \gamma^2}{8\pi^2 \sqrt{\sin^2 x + \sin^2 y}} e^{i(x (n-j)+y(k-l))}\, ,
\end{equation}
where here we used subindices $(i,j)$ to denote coordinates $x,y$ in the square lattice.

The idea is then to consider fixed shapes in the lattice and evaluate $g(A,B)$ as we increase the number of points. Performing inverse-power-law fits to the data ---\eg $\{ 1/x^3,1/x^2,1/x,1\}$ --- we can read off the continuum results from the $\mathcal{O}(1)$ coefficients. In Table \ref{tbl1} we present the results obtained for identical pairs of disks, squares, and rectangles of side-lengths: $2L\times L$, $4L\times L$ and $6L\times L$, respectively. As we can see, the numerical result obtained for the disks is very close to the analytic one appearing in \req{gdisksf}. Interestingly, the result for a pair of squares is also remarkably similar to the disks result. However, as we deform the squares and replace them with increasingly thinner rectangles, the coefficient decreases, clearly differing from the disks result. 

\begin{table}[t] 
  \centering
 \begin{tabular}{|c ||c| c |c| c| c| } 
 \hline
  &  \small  $\quad$ Disks $\quad$ &  \small $\,\,$  Squares $\,\,$ & \small   Rect. $(2L\times L)$ &   \small  Rect. $(4L\times L)$ & \small Rect. $(6L\times L)$  \\ 
 \hline\hline
 $g(A,B) \times \frac{15 \pi^2}{16}$& $0.994$ & $0.997$  & $0.938$ & $0.862 $ & $0.840$ \\ 
 \hline
\end{tabular}
 \caption{Continuum values of the coefficient  $g(A,B)$ normalized by the analytic result corresponding to a pair of disks, $16/(15\pi ^2)$, obtained in the lattice for different pairs of identical regions.}
\label{tbl1}   
\end{table} 

In fact, the limit of very thin parallel plates with area ${\cal A}$ can be treated analytically in any dimension. In this case, we have a limit of translational invariance in the direction parallel to the plates. Consequently, the eigenvectors of the correlator decompose as
\be
u^{\lambda}_{A}(x)\sim {\cal A}^{-1/2}\, e^{i \vec{k}_\parallel\vec{x}_\parallel} \phi_{\lambda,\vec{k}_\parallel}(x_\perp) \,. 
\ee
Upon averaging over the region all vectors with $\vec{k}_\parallel\neq 0$ do not contribute. Only the constant eigenvectors in the parallel directions contribute. On these, the correlator has the same effect as the correlator of a $d=2$ dimensional field, since the integration imposes zero parallel momentum in the parallel direction,
\be
\int \diff^{d-2}x_\parallel C(\vec{x})= \int \diff^{d-2}x_\parallel \int \frac{\diff^d p}{(2\pi)^d} \frac{\,e^{-i p x}}{i p\hspace{-.2cm}\slash}\gamma^0=  \int \frac{\diff^2 p}{(2\pi)^2} \frac{\,e^{-i p x}}{i p\hspace{-.2cm}\slash}\gamma^0= C_2(x^1)\,,  
\ee
where, however, the spinorial representation is still of dimension $2^{[\frac d2]}$. The problem then collapses to a two dimensional problem of the mutual information of $2^{[\frac d2]-1}$  fermions for well separated intervals. Each $d=2$ fermion has $g(A,B)=1/3$. We then obtain from (\ref{gbbs})
\be
g(A,B)|_{\textrm{plates}} =  \frac{\(\Gamma\!\(\frac d2\)\)^2 2^{[\frac d2]-1}}{3 \pi^{d-2}}\,.
\ee
Therefore, in any dimensions we have 
\be
\frac{g(A,B)|_{\textrm{plates}}}{g(A,B)|_{\textrm{spheres}}}= \frac{\pi \Gamma\!\(\frac d2\)\,\Gamma\!\(d+\frac12\) }{3 \, 2^{d-1}\,\(\Gamma\!\(\frac{d+1}2\)\)^3} \,.\label{number}
\ee
To compare with the numerical calculation above for rectangles, this gives for $d=3$ a ratio of $\sim 0.771$. The results in Table \ref{tbl1} asymptotically approach this value as the rectangles become thinner. The ratio (\ref{number}) is always less than $1$ and decreases with dimension.

Hence, we conclude that the long-distance limit of the free fermion mutual information differs from the EMI one, as the latter depends on the geometry of the entangling regions exclusively through the product of volumes, whereas the former includes an additional shape-dependent function.

\section{Is the EMI the mutual information of a CFT?}\label{isemicft}

We have seen that the conjectural EMI theory must contain a fermion with dimension $(d-1)/2$ as a lowest-dimensional operator. We have also shown that the long-distance leading term of the free fermion and EMI models do not coincide, while the EMI gives exactly the leading conformal block to the MI of the free fermion for spheres. We now show this is not compatible with the EMI model being the MI of a fixed CFT.   

First, recall that if a QFT contains a field $\psi$ with the two-point function of a free field, then this field has to be free, that is, all its correlators satisfy Wick's theorem \cite{streater2000pct}. The key of the argument is that the two-point correlator satisfies a local equation. In consequence, the field operator itself must satisfy the free equation of motion. For example, for a field having the two-point function of the free fermion field, writing $u(x)=\gamma^\mu \partial_\mu \psi(x)$, we get $\langle 0|u^\dagger(x) u(y) |0\rangle=0$. From this, the smeared operator $u_\alpha=\int \diff x\, \alpha(x) u(x)$, satisfies $|u_\alpha |0\rangle|^2=0 \implies u_\alpha |0\rangle=0$, and since no local operator can annihilate the vacuum,  $\gamma^\mu \partial_\mu \psi(x)=u(x)=0$. For a field satisfying the free equation of motion, the usual decomposition into positive and negative frequencies can be done, and the usual numeric commutator (anti-commutator) follows \cite{streater2000pct}. The c-number commutator implies Wick's theorem. This determines all correlators of the field with itself. Further, in the Hilbert space generated by acting with the field polynomials on the vacuum, the only other field operators that are mutually local (commuting or anticommuting at spacelike distances) with $\psi$ are polynomials on $\psi$ and its derivatives \cite{epstein1963borchers}. This means that the free field decouples from the rest of the theory and we have a tensor product of theories and Hilbert spaces 
\be
{\cal T}_{\ssc \textrm{EMI}}={\cal T}_1\otimes {\cal T}_\psi\,,\hspace{1cm}{\cal H}_{\ssc \textrm{EMI}}={\cal H}_1\otimes {\cal H}_\psi\,.      \label{fff}    
\ee

As a consequence of this decoupling, it is not difficult to see that the EMI mutual information cannot correspond to a CFT. First, the leading long-distance contribution for any shapes is dominated by the leading primary operator contribution to the OPE of the R\'enyi operators, which here is a free fermion field. The coefficient of this contribution depends on the modular flow of the regions acting on the fermion field. This modular flow is not universal for non-spherical regions and depends on the operator content of the full theory. However, as the fermion field decouples, the modular flow acting on the fermion field has to be the same as the flow of the free fermion theory. This gives the same coefficient for the long-distance contribution for the EMI and the free fermion for any shapes. This contradicts the results of section \ref{longdist}. 

Another important consequence of this decoupling is that \req{fff} would imply
\be
I_{\ssc \textrm{EMI}}(A,B)=I_1(A,B)+I_{\ssc \psi}(A,B)\,, \label{ggg11}
\ee
and therefore
\be
I_{\ssc \textrm{EMI}}(A,B)\ge I_\psi(A,B) \label{ty}
\ee
for any $A$, $B$. The coefficient of the EMI model has to be adjusted to match the long-distance contribution of the free fermion for balls since $I_1(A,B)$ falls to zero faster at large distances. However, with this calibration, (\ref{ty}) is not violated for the contributions to the mutual information we have studied so far. Then, it does not give a different argument for the impossibility of the EMI model. 

Another startling feature of EMI as a MI of a CFT is that for two spheres it consists entirely of a single conformal block of spin $1$ and dimension $d-1$. This same feature holds for a free fermion in $d=2$. In that case, it is the result of a precise cancellation between the contributions of different primary fields in the replicated theory. We turn now to analyze whether this is possible for $d > 2$.   

\section{Can EMI represent a limit of CFTs?}\label{emilimits}

No fixed QFT can have a MI described by the EMI model for $d>2$. However, this result by itself does not imply that the EMI model has no physical meaning. To see this in perspective, we could think of applying the same ideas to extract the QFT from the mutual information for holographic EE \cite{ryu2006holographic}. The bulk minimal area in AdS is known to provide a formula for the entropy whose MI satisfies all the axioms (\ref{pos}-\ref{area}), plus conformal invariance. However, if we try to bootstrap the holographic formula to obtain information on the operator content, we immediately encounter a problem: the long-distance MI is trivial since there are phase transitions that make it vanish for a long enough distance. A naive application of the same principles we have applied to the EMI model would lead to the conclusion that there are no operators in the theory, and that the Ryu-Takayanagi formula is inconsistent. Of course, as it is well known, the area term represents only the leading $N^2$ term in the holographic entropy, and there are subleading $N^0$ terms that do not vanish at large distances. They are given by the mutual information of free fields in the bulk entangling wedges \cite{Faulkner:2013ana}. Therefore, from the knowledge of the Ryu-Takayanagi term alone we cannot infer the physical validity of the model as the EE of a QFT comparing with the expected result for a fixed QFT.      

This highlights a difference in the nature of the analysis of the validity of an entropy (or mutual information) function as compared to, \eg  checking the validity of a system of correlation functions. The reason is precisely that entropy is a quantity that can be defined universally for any QFT. There are several ways in which an entropy function satisfying (\ref{pos}-\ref{area}) can be obtained from valid QFTs but which do not represent the entropies of any QFT. In particular, the system of requirements (\ref{pos}-\ref{area}) do not fix a normalization of the entropy, and this is related to the fact that any linear combination with positive coefficients of entropies of two theories will also satisfy the requirements. This is justified by the fact that entropies for a tensor product of theories are the sum of the corresponding entropies. Moreover, if we have a set of QFTs determined by a parameter, we can take limits on this parameter and the limit of the entropy will also satisfy the requirements. This parameter may be a coupling constant or some other parameter like the number $N$ of colors, as in large-$N$ models. Therefore, there may be many solutions to the constraints that do not represent actual QFTs but specific limits of QFTs. The holographic EE is one of these limits. 

A natural question is if the EMI model represents some limit of CFTs. The main argument of the preceding section relies on the coefficient of long-distance mutual information for regions of arbitrary shape. These coefficients depend on the modular flow of the theory, which is not universal for nonspherical regions and depends on the full operator content. Then, with our present rather incomplete knowledge about modular Hamiltonians, it is not clear that the argument still holds for limits of theories. In taking this limit, we cannot discard from the outset having a set of fermions taking a free limit while the modular flow producing significant persistent differences in the coefficient with respect to the decoupled free fermion field.   
        
However, we will answer the question of whether or not the EMI is a limit of QFT's in the negative. To do so, we turn attention again to spheres, where the modular flow is universal.

\subsection{Further analysis of the free fermion MI for spheres}
As the EMI contains a free fermion (as a limit), it must contain other contributions of the same field. In the long-distance limit, the leading such contribution comes, in the replica trick, from the field $\psi$ in one copy and the derivative $\partial \psi$ in another copy. Let us determine this operator with more precision. 

The OPE of a twist operator for the R\'enyi entropy of integer order $n$ for a sphere is of the form
\be
\tau^{(n)}=\sum C^{(n)}_\alpha\, {\cal O}^\alpha\,, 
\ee
where the sum is over all local operators in the $n$-replicated theory, $\alpha$ are Lorentz indices, and the coefficients depend only on the theory and the geometric features of the sphere. Namely, the tensor structure of $C^{(n)}_\alpha$ can only depend on the time-like unit vector $n_\mu$ determining the orientation of the sphere in space-time, and $g_{\mu\nu}$. The leading long-distance term comes from the two copy operator (to simplify the notation we call $1,2$ the two copies, which can be any pair of copies of the $n$ replicas)
\be
V^\mu=\bar{\psi_1} \gamma^\mu \psi_2-\bar{\psi_2} \gamma^\mu \psi_1\,.   \label{calc}    
\ee
 This gives a contribution $\sim n_\mu V^\mu$ to $\tau^{(n)}$. 
The minus sign in (\ref{calc}) comes from charge-conjugation invariance. Other Dirac matrix structures do not contribute because of parity invariance or because they form antisymmetric tensors whose contraction with products of $n^\mu$ or the metric vanishes \cite{casini2021mutual,chen2017mutual}.

The above operator has dimension $d-1$. The next operator comes from adding a derivative to the fermion bilinear, with dimension $d$. The descendant field
\be
 \partial_\alpha V^\mu=(\partial_\alpha\bar{\psi_1}) \gamma^\mu \psi_2+\bar{\psi_1} \gamma^\mu \partial_\alpha\psi_2- (\partial_\alpha\bar{\psi_2}) \gamma^\mu \psi_1- \bar{\psi_2} \gamma^\mu \partial_\alpha\psi_1  \,,   
\ee
is already included in the first conformal block studied in section \ref{confblock}. We recall that this conformal block contribution to the mutual information exactly coincides with the EMI for spheres. Then, we consider other bilinear operators with a single derivative that forms independent primary operators. Again, charge conjugation, parity, and the fact that this operator cannot have an antisymmetric structure to give a non-vanishing contribution leaves as the only possibility the two-copy operator 
\be
V_{\alpha\mu}=\frac{1}{2}\left[(\partial_\alpha\bar{\psi_1}) \gamma_\mu \psi_2-\bar{\psi_1} \gamma_\mu \partial_\alpha\psi_2+ (\partial_\alpha\bar{\psi_2}) \gamma_\mu \psi_1- \bar{\psi_2} \gamma_\mu \partial_\alpha\psi_1+(\alpha\leftrightarrow \mu)\right]\,. \label{ffff}
\ee
Notice this operator is traceless due to the Dirac equation. It is a primary field of spin $2$ (symmetric traceless two index tensor) of dimension $d$. For the free fermion, no other contribution of dimension less or equal to $d$ is possible if $d>2$. For $d=2$ we have the contribution of four fermions in four copies, with dimension $2(d-1)=2$. In $d=2$ this contribution exactly cancels the one from (\ref{ffff}) because the full result coincides with the first conformal block. This will not be possible for $d> 2$. 

The two-point function has the form\footnote{The general structure of correlators of primary fields in a CFT consists of tensor products of the $I$ tensor projected on the adequate symmetry representation. Only one projector is enough in (\ref{correla}) since $P$ commutes with the tensor product of the $I$.   }
\be
\langle V_{\mu\nu}(0) V_{\alpha\beta}(x)\rangle\propto  \, |x|^{-2d}\, P_{\mu\nu,\delta_1\delta_2} I^{\delta_1 \gamma_1}(\hat{x}) I^{\delta_2\gamma_2}(\hat{x})  \label{correla} P_{\gamma_1\gamma_2,\alpha\beta}\,,
\ee
where $P=P^2$ is the projector on the symmetric traceless tensors
\be
P_{\mu_1\mu_2,\nu_1\nu_2}=\frac{1}{2}\left( g_{\mu_1\nu_1} g_{\mu_2\nu_2}+g_{\mu_1\nu_2} g_{\mu_2\nu_1}-\frac{2}{d} \,g_{\mu_1\mu_2} g_{\nu_1\nu_2}\right)\,,
\ee
and $I$, $I^2=1$, is the tensor
\be
I_{\mu\nu}(\hat{x})=g_{\mu\nu}-2 \hat{x}_\mu \hat{x}_\nu\,.
\ee

The contribution to the R\'enyi operators for each sphere is then proportional to operators of the form $n^\mu n^\nu V_{\mu\nu}$, summed over all possible pairs of copies. The contribution to the mutual information to the lowest order has the same tensor structure of the correlation between these operators  
\be
\Delta I(A,B)\propto  n_A^\mu n_A^\nu\, \langle V_{\mu\nu}(0) V_{\alpha\beta}(L \,\hat{l})\rangle n_B^{\alpha} n_B^\beta\,.
\ee
The coefficient can be computed by the method of \cite{casini2021mutual}. We relegate to appendix \ref{coeff} the details of this calculation. We get
\be
\Delta I(A,B)= - \frac{ \sqrt{\pi}\, \Gamma\!\(d+1\) 2^{[\frac d2]-1}}{  \Gamma\(d+\frac 32\)}\,\frac{R_A^d R_B^d}{L^{2 d}}\, \left[(2(n_A\cdot l)(n_B\cdot l)- (n_A\cdot n_B))^2 -\frac{1}{d}\]\,.\label{hh}
\ee

Let us analyze whether this contribution to the free fermion can be canceled by some other contribution coming from additional fields. The lowest-dimensional contribution by a primary field of dimension $\Delta$ of the theory, different from the fermion, comes from the field in two copies and gives a dependence $L^{-4 \Delta}$. Then we must have $\Delta\le d/2$. Because of the unitarity bound, this leaves us with a scalar, a spin $1/2$ fermion, or a helicity one field (a completely antisymmetric field of $d/2$ indices). This last possibility only exists for even dimensions, and, for $\Delta=d/2$ saturates the unitarity bound. Then it must be a free field. 

In fact, the tensor structure of the contribution (\ref{hh}) exactly matches the leading term of a helicity-$1$ field in any even dimensions and does not match the scalar or fermion contribution \cite{casini2021mutual}. This eliminates the scalar and fermion as possibilities. However, in odd dimensions, there is no such conformal primary of dimension $d/2$ and helicity equal to $1$. The only possibility to eliminate this contribution from the free fermion in the EMI model in odd dimensions would be to obtain it from descendant fields of a scalar. The contribution of the derivative of a free scalar could be present without the contribution of the scalar itself if the theory is generated by polynomials of $\partial_\mu \phi$. In this case, the theory is not conformally invariant and only scale invariant. The derivatives of the free scalar form the subalgebra stable under $\phi\rightarrow \phi+ \textrm{const}$. The contributions for spheres can be analyzed with the same methods. The contribution coming from fields $\partial_\mu \phi_1 \partial_\nu \phi_2$ in two copies does not have the correct tensor structure. This is because, apart from a contribution independent from the orientation of the spheres, produced by $\partial_\mu \phi_1 \partial^\mu \phi_2$, the tensor structure must be of the form
\be
n_A^\mu n_A^\nu \,\langle \partial_\mu\phi_1(r_A)\partial_\nu \phi_2(r_A) \partial_\alpha\phi_1(r_B)\partial_\beta\phi_2(r_B)\rangle\, n_B^\alpha n_B^\beta\sim  \frac{1}{L^{2 d}}\, (d\, (n_A\cdot l)(n_B\cdot l)-(n_A\cdot n_B))^2\,.
\ee
 This eliminates the possibility that the EMI can be produced by a limit of QFTs in odd dimensions.

For even dimensions, as the term (\ref{hh}) comes with a negative coefficient, we can cancel it by the contribution, with the adequate proportion,  of a free helicity $h=1$ field, which has the same tensor structure \cite{casini2021mutual}
\be
I_{h=1}(A,B)\sim  \frac{\sqrt{\pi} \, \Gamma\!\(d+1\) \Gamma\!\(d-1\)}{\Gamma\!\(d+\frac32\) \(\Gamma\!\(\frac d2\)\)^2}\frac{R_A^d R_B^d}{L^{2 d}}\, \left[(2(n_A\cdot l)(n_B\cdot l)- (n_A\cdot n_B))^2 -\frac{1}{d}\right]\,,\label{h1}
\ee  
This would imply    
\be
I_{\ssc \rm EMI}=  \,I_{\ssc \textrm{fermion}}+ b \, I_{h=1}+\cdots\,,\label{prol}
\ee
where $b$ is the ratio (with positive sign) between the coefficients of (\ref{hh}) and (\ref{h1}),
\be
b= 2^{[\frac d2]-1} \frac{\(\Gamma\!\(\frac d2\)\)^2}{\Gamma\!\(d-1\)}  \,.
\ee

In (\ref{prol}) we have again normalized the EMI to have the long-distance MI of the single Dirac fermion for spheres. 
Now, as the ellipsis in (\ref{prol}) represent other positive contributions to the MI of the spheres, we can check what happens for nearly-touching spheres, where we can extract the coefficient $k$ of the area term.  From (\ref{prol}) we should have
\be
k_{\ssc \rm EMI}\ge k_{\ssc \textrm{fermion}} +b \, k_{h=1} \,.\label{docee} 
\ee
For $d=4$, using the results of section (\ref{not}), and the ones for a Maxwell field $k_{\ssc \rm Maxwell}=0.0110$ \cite{Casini:2015dsg}, we get a left hand side $0.0273$ and for the right hand side $0.0325$. This is incompatible with (\ref{docee}). For higher even dimensions  the number of field components in the totally antisymmetric field of $d/2$ indices (the $h=1$ field) is $\left(\begin{array}{c} d \\ d/2  \end{array}\right)$, from which $2\left(\begin{array}{c} d-1 \\ d/2+1  \end{array}\right)$ are constraints. Half the rest of the components give momentum variables in the Hamiltonian formalism. The value of the $k$ for this field is then (see the similar calculation for a Maxwell field in \cite{Casini:2015dsg})
 \be
 k_{h=1}= \frac{1}{2}\left(\left(\begin{array}{c} d \\ d/2  \end{array}\right)- 2\left(\begin{array}{c} d-1 \\ d/2+1  \end{array}\right)\right) \, k_{\ssc \rm scalar}\,. 
 \ee 
For dimension $\ge 6$ we can use the very good approximations \cite{Casini:2009sr}
\be
 k_{\ssc \rm scalar}\sim \frac{\Gamma\!\(\frac{d-2}2\)}{2^{d+2} \pi^{(d-2)/2}}\,,\hspace{.6cm} k_{\ssc\rm fermion}= 2^{[\frac d2]}\, k_{\ssc \rm scalar}\,.
\ee
Plugging this into (\ref{docee}) we see that for $d=6$ the left hand side is $0.00438$ while the right hand side is $0.00459$. Then the inequality does not hold for $d=6$. However, it still holds for higher dimensions. This shows the mismatch between the EMI and the fermion for spheres cannot be remedied by adding other fields for $d=6$. For higher even dimensions the analysis should be extended further in the MI expansion. We will not attempt it here.  

In conclusion, there is no way in which the EMI result could be the mutual information for a CFT, even considered as a limit of theories. This is because this limit of theories must contain a field (or many fields) with the two-point function of the free fermion. In that case, the next-to-leading-order term (\ref{hh}) for the mutual information must also be present, disregarding the particular limit of theories, because it depends only on the two-point function and the universal form of the modular flow for spheres. The EMI does not contain such a term, and no such term could be canceled by other possible contributions.

\section{Discussion}\label{discu}

In this paper, we have shown that the EMI model is not compatible with the mutual information of a QFT (for $d>2$). This means that $I_3\equiv 0$ is inconsistent in relativistic QFT for $d>2$. Moreover, this shows that the known set of constraints for the mutual information in QFT is incomplete. Below we make some further comments on this result interpreted in terms of twist operators, the relation with the pinching property of the MI of free fields, and with reflection positivity inequalities satisfied by correlation functions.

\subsubsection*{Coherent twist operators}
In the task of constructing viable entropy functions, a natural route would be to have an ansatz for the behavior of the R\'enyi twist operators. The simplest candidate would be to produce them out of free fields. This is, in fact, the way in which formulas analogous to the EMI model were proposed in \cite{swingle2010mutual}. Suppose we conjecture twist operators with the form of coherent operators
\be
\tau_A^{(n)}=e^{i \int_A \diff x\, \alpha^{(n)}_A(x)\, \phi^{(n)}(x)}\,, \label{ssc}
\ee
where $\phi^{(n)}(x)$ is some generic Gaussian field (free or generalized free field), and $\alpha^{(n)}_A(x)$ is a function with support in the causal region $A$ and determined by the geometry of $A$. For example, it may have support only in the boundary of $A$, or in any Cauchy surface of $A$. In this latter case, it has to be conserved for different Cauchy surfaces. Under this assumption, we get for the mutual information a bilinear expression
\be
I(A,B)=\lim_{n\rightarrow 1}(n-1)^{-1}\log\left(\frac{\langle\tau_A^{(n)} \tau_B^{(n)}\rangle}{\langle \tau_A^{(n)}\rangle\langle \tau_B^{(n)}\rangle}\right)=  \int_A \diff x\, \int_B \diff y\, \alpha_A(x)\, \alpha_B(y)\, G(x-y)  \, ,\label{G}
\ee
for some geometric functions $\alpha_A, \alpha_B$, and where $G(x-y)$ is some two-point function. This form of the entropy immediately gives $I_3(A,B,C)=0$ for disjoint regions. Hence, this property seems naturally associated to twist operators which are exponentials of Gaussian fields. 

However, in most cases for the functions $\alpha_A^{(n)}$ and the free fields $\phi^{(n)}$, the result for $I(A,B)$ is not monotonic. In fact, the only monotonic possibility for a MI with the structure (\ref{G}) is the EMI function \cite{Casini:2008wt}, with the precise form (\ref{aa}) or equivalently (\ref{sis2}). In consequence, monotonicity imposes to the ansatz (\ref{ssc}) a precise form, where the twist operators are exponentials of the flux of a current (local charge operator)
\be
\tau_A^{(n)}\sim e^{i\int_{\Sigma_A} \diff \sigma\, J^{(n)}_\mu \,\eta_A^\mu} \,,  \label{saga}
\ee
but where the conserved current operator $J^{(n)}_\mu$ must be Gaussian. For $d=2$ this is the case of the Dirac current by bosonization. For $d>2$, this implies $J^{(n)}_\mu$ is a generalized free field (GFF).  It is direct that evaluating the logarithm of the expectation values of these type of twist operators we get the expression (\ref{aa}) for the MI of the EMI model,
\be
I_{\rm \ssc EMI}(A,B)\propto  \left< \left(\int_{\Sigma_A} \diff \sigma\, J_\mu \,\eta_A^\mu\right) \left(\int_{\Sigma_B} \diff \sigma\, J_\mu \,\eta_B^\mu\right) \right>\,.\label{nece}
\ee 

This has an interesting consequence. If we inquire whether the EMI model determines a QFT the answer is yes, but with the expression of $I_{\rm \ssc EMI}(A,B)$ not corresponding to the mutual information, but rather to a correlation function of fluxes of GFF current. The Hilbert space and operator content can then be recovered in the usual way a QFT can be recovered from the correlation functions, and the result is the theory of the GFF current.  

However, we have seen that the EMI cannot be interpreted as a MI for $d>2$. Recovering a QFT from the mutual information is very different (and more uncertain) than recovering it from correlation functions. For $d=2$, a free current can be recovered from the EMI interpreted as correlators, but, presumably, the fermion theory should be recovered interpreting it as a MI. The MI of a theory of a free current is a very different function from the one of the fermion \cite{arias2018entropy}. 

Therefore, the R\'enyi twist operators cannot be coherent operators for $d> 2$. For a free fermion theory, the twist fields can indeed be expressed as exponentials of fluxes of currents \cite{casini2005entanglement},
\be
\tau^{(n)}_A= e^{i \int_{\Sigma_A} \diff \sigma\, \eta_\mu \left(\sum_{k=-(n-1)/2}^{(n-1)/2} \frac{2\pi k}{n} J_{(k)}^\mu \right)}\,,
\ee
where $J_{(k)}^\mu$ are mutually independent currents for free Dirac fields in the replica space. These currents are however not free fields for $d>2$; the expectation values of these operators give determinants and not exponentials of quadratic forms. This can be seen as the origin of the difference between the free fermion and the EMI model for $d>2$.  

\subsubsection*{Pinching property}
The expression (\ref{saga}) for the twist operators, with $J^{(n)}_\mu$ a generalized free field current, clashes directly with another expected property of the entropy. This is called the pinching property \cite{casini2021mutual}. We have not listed it among the general properties of the entropy in the introduction because it expresses a relation of a property of the MI with properties of the QFT, rather than constraints on the MI function itself. This property is as follows. 

Consider the null future horizon $H$ of the causal set $A$ corresponding to a sphere of radius $R$. This null surface is parametrized by the angular direction $\vec{\Omega}$ determining a null ray, and an affine parameter $s\in(0,R)$ along the null segments. We take $s=R$ as the position of the spherical base of the null surface and $s=0$ as the tip of the cone. Let us consider the causal region $A(\delta,\epsilon)$ determined by its future horizon: 
\be
H-\left\{(\vec{\Omega},s)\,\slash\, |\vec{\Omega}-\vec{\Omega}_0|<\epsilon, s> \delta \right\}  \,.
\ee 
This results from cutting from $H$ a segment of a pencil of null generator around a fixed, but arbitrary, direction $\vec{\Omega}_0$. 
In the limit $\delta\rightarrow 0$, that is, when the cut on the null surface reaches the tip of the cone,  the region $A(\delta,\epsilon)$ collapses to a null surface, and the causal region does not have any space-time volume. In the limit $\epsilon\rightarrow 0$, the region converges to the full spherical region. We consider the mutual information $I(A(\delta,\epsilon), B)$ for any fixed region $B$. The question is about the limit $\delta,\epsilon\rightarrow 0$ of this mutual information. 

For interacting CFTs, including GFF, the correlation functions are too singular to obtain operators when smearing field operators on a null surface. Then the limit $\delta\rightarrow 0$ eliminates the algebra, and we have the pinching property \cite{casini2021mutual}
\be
\lim_{\epsilon\rightarrow 0} \lim_{\delta\rightarrow 0} I(A(\delta,\epsilon), B)=0\,.
\ee
For a free theory ---in the sense of being the theory of a field satisfying a local linear equation of motion, not a GFF--- the algebra of operators can be localized on the null surface, and we have
\be
\lim_{\epsilon\rightarrow 0} \lim_{\delta\rightarrow 0} I(A(\delta,\epsilon), B)= I(A,B)\,.\label{b}
\ee
When an interacting theory has a free sector,  the limit is non zero, but still 
\be
\lim_{\epsilon\rightarrow 0} \lim_{\delta\rightarrow 0} I(A(\delta,\epsilon), B)\lneq I(A,B)\,.  
\ee

 It is easy to see that the EMI model gives a non-pinching entropy function, satisfying \req{b} just like the MI of free fields because the EMI is an integral over the boundary of the regions.  The MI of a replicated theory is just the number of replicas times the mutual information of the theory. According to the ansatz (\ref{saga}) for the description of the twist operators of the EMI model, the algebra of this replicated EMI theory would contain a generalized free current. This must lead to discontinuities under pinching for this replicated model and hence contradicting the form of the mutual information of the EMI model. The only allowed operator content for continuity under pinching are free fields and their Wick polynomials.     
 
\subsubsection*{New entropy inequalities?}

In this paper, we showed that the known properties of the entropy in QFT are not enough to allow for a complete description of the space of possible entropy functions. An important question is then what further properties are missing. A natural surmise is that there should be additional unknown inequalities. Indeed, operator correlations obey infinitely many inequalities and are precisely these inequalities the ones that allow for the reconstruction of the Hilbert space scalar product and the quantum theory. The R\'enyi entropies of integer order $n\ge 2$, as they are produced by expectation values of twist operators, also satisfy this infinite set of inequalities. In real time, these operator inequalities ---equivalent to the reflection positive inequalities in imaginary time--- are called wedge reflection positivity, Rindler positivity, or CRT positivity (for charge-reflection-time inversion) \cite{casini2010entropy,casini2012positivity,blanco2019Renyi}. 

For the R\'enyi entropies, taking a series of regions $A_i$, $i=1,\cdots,m$, contained in the right wedge $x^1-|x^0|>0$, and defining the reflection $\bar{x}=(-x^0,-x^1,x^2,\cdots,x^{d-1})$ carrying the right wedge into the left one, we have     
\be
\det \left(\lbrace e^{(n-1) I_n(A_i,\bar{A}_j)}\rbrace_{i,j=1}^m\right)\ge 0\,. \label{det1}
\ee
The $I_n(A,B)=S_n(A)+S_n(B)-S_n(AB)$ are the R\'enyi mutual informations. If we take all regions $A_i$ to be equal except for infinitesimal differences, we get an infinite series of inequalities involving derivatives of the R\'enyi entropies of arbitrarily large order.  
If we naively think in taking the limit $n\rightarrow 1$ in these inequalities we get an infinite set of polynomial inequalities for the mutual information 
\be
\det\left(\lbrace I(A_i, \bar{A}_j)  + I(A_{i+1}, \bar{A}_{j+1})-I(A_i, \bar{A}_{j+1})-I(A_{i+1},\bar{A}_j) \rbrace_{i,j=1,\cdots,m-1}\right)\ge 0\,.\label{det}
\ee   
However, this set of inequalities for the entropy, does not hold in general theories and regions. An example is the holographic EE \cite{casini2012positivity}. The reason is simply that it is not possible to analytically continue the inequalities down to $n=1$. 

The present investigation suggests that inequalities generalizing the reflection positivity inequalities for operators is not the right idea to get the correct additional constraints. In fact, somewhat ironically, the EMI model, which obeys all the known constraints for a MI in QFT, and is not the MI of any QFT, also obeys the infinite series of inequalities (\ref{det}). The reason is that the EMI comes from the expectation value of operators, as shown by \req{nece}. 

A natural expectation is that there are more inequalities following a different line, and generalizing strong subadditivity (SSA) rather than operator inequalities. In this same direction, it has recently been shown that SSA leads to unitarity bounds for fields with the tensor structure of free conformal primaries \cite{casini2021mutual}. In this case, certain SSA inequalities are saturated for the conformal dimension at which the field becomes free.  For fields with other tensor structures, it may well be the case that other inequalities generalizing SSA are needed to do the job of producing the unitarity bounds. 

\section*{Acknowledgments}
H.C. is indebted to Gonzalo Torroba for many useful insights on the subject of CFTs. 
This work was supported by the Simons foundation through the It From Qubit Simons collaboration.
H.C. was also supported by CONICET, CNEA, and Universidad Nacional de Cuyo, Argentina.

\appendix
\section{EE universal coefficients for spheres and strips for the EMI } \label{stripsphere}

From the results presented in section \ref{confblock}, we can extract two universal characteristic coefficients appearing in the entanglement entropy of the EMI model for general dimensions. These correspond, respectively, to the ``strip coefficient'' ---characterizing entangling regions which are very large in all directions but one--- and the one corresponding to spherical entangling surfaces. 

%\subsection{Strip coefficient}
Whenever we have an entangling region $A$ which is large in all directions, with a total area of $\mathcal{A}$ in such directions, and a small transversal dimension of length $r$, the entanglement entropy universal term takes the form
\begin{equation}
\see |_{\rm univ}= - k^{(d)} \frac{\mathcal{A}}{r^{d-1}}+\dots
\end{equation} 
where the dots denote subleading contributions.
This coefficient is known analytically in general dimensions \eg for holographic theories dual to Einstein gravity \cite{Ryu:2006ef} and  for free scalars and fermions \cite{Casini:2009sr} in terms of the two-dimensional entropic c-function $c(L)\equiv L\cdot (\diff \see/\diff L)$. 

The coefficient $k^{(d)}$ can be extracted using the mutual information as follows. The idea is to consider two parallel entangling regions with the same shape as the one corresponding to the large dimensions of $A$ in the previous paragraph, separated a distance $r$. In that case, we have
\begin{equation}\label{mutuuu}
I = k^{(d)}\frac{ \mathcal{A}}{r^{d-1}}+\dots 
\end{equation}

For the EMI model, we can then exploit the calculation for two concentric spheres presented in section \ref{confblock} to obtain $k^{(d)}$. In order to do that, we take the final expression in \req{EMI-zero-boost} for unboosted spheres and rewrite it in terms of the radii $R_A$, $R_B$. Then, we consider
\begin{equation}
R_A= R-\delta/2\, , \quad R_B=R+\delta/2\, , \quad \delta \ll R\, , 
\end{equation}
which puts the spheres very close to each other. Then, the result takes the form expected from \req{mutuuu},
\begin{equation}\label{sksk}
I = k^{(d)}\frac{ \mathcal{A}(\mathbb{S}^{d-2})  }{\delta^{d-1}}+\dots  \quad  \text{where}\quad \mathcal{A}(\mathbb{S}^{d-2})  \equiv \frac{2 \pi^{\frac{d-1}{2}} R^{d-2}}{\Gamma\!\(\frac{d-1}{2}\)}\, ,
\end{equation}
is the area of the $\mathbb{S}^{d-2}$. Doing this, we find for the EMI
\begin{equation}\label{kemid}
k^{(d)}_{\rm \ssc EMI}=\frac{ 2 \pi^{\frac{d-2}{2}} \Gamma\!\(\frac{d-2}{2}\)}{\Gamma\!\(d-2\)} \kappa_{(d)} \, .
\end{equation}

%\subsection{Sphere coefficient}
When putting the two concentric spheres very close together, we are in practice using mutual information as a geometric regulator of entanglement entropy ---see \cite{Casini:2006ws,casini2015mutual} for more on this approach. Hence, in addition to the leading area-law-like piece in \req{sksk}, we can also extract the universal characteristic coefficients appearing in the logarithmic (even $d$) and constant (odd $d$) coefficients. In general we will have\footnote{Our conventions agree with those of \cite{Myers:2010xs}, for instance.}
\begin{equation}\label{ikeo}
I = k^{(d)}\frac{ \mathcal{A}}{r^{d-1}}+\dots +2 \left\{ \begin{array}{lr} (-1)^{\frac{d}{2}-1} 4 A^{(d)}\log(R/\delta) & \text{(even $d$)\, ,} \\ (-1)^{\frac{d-1}{2}} F^{(d)}  & \text{(odd $d$)\, .}  \end{array}\right.
\end{equation}
In this expression, $A$ coincides with the trace-anomaly coefficient controlling the Euler density $E_d$ for  even dimensional theories \cite{Duff:1993wm}
\begin{equation}\label{traceano}
\braket{T^a_a}= \sum_i B_i I_i  - (-1)^{\frac d2} 2A^{(d)} E_d \, ,
\end{equation}
and $  (-1)^{\frac{d-1}{2}} F^{(d)}$ equals the free energy of the corresponding CFT when put on a round $\mathbb{S}^d$ \cite{Casini:2011kv}.

For the EMI model, the expansion \req{sksk} contains the terms expected from \req{ikeo} and we identify
\begin{equation}\label{FAemi}
F_{\rm \ssc EMI}^{(d)}= \frac{2\pi^{d-1}}{\Gamma\!\(d-2\)} \kappa_{(d)}\, , \quad A_{\rm \ssc EMI}^{(d)}=\frac{\pi^{d-2}}{\Gamma\!\(d-2\)} \kappa_{(d)}\, .
\end{equation}

\section{OPE blocks and the EMI\label{B}}

In this appendix, we are interested in studying the conformal block associated to a conserved current $J^\mu$ with  $\Delta=d-1$ and $J=1$ and find in this the EMI formula for disjoint spheres. One can find parts of the material presented in this appendix in section 2 of \cite{Czech:2016xec} and section 3 of \cite{deBoer:2016pqk}. For a recent review of the shadow operator formalism and the original literature, see \cite{SimmonsDuffin:2012uy} and references therein. 

In conformal field theories the set of quasiprimary operators and their descendents represents a complete basis of operators.  Thus, when other operators are located away from points $x$ and $0$ the product of say $\mathcal{O}_i(x)$ and $\mathcal{O}_j(0)$ inside an arbitrary correlator can be expressed as 
\bea\label{OPE-expansion}
\mathcal{O}_i(x)\mathcal{O}_j(0)=\sum_{k}C_{ijk} |x|^{\Delta_k-\Delta_i-\Delta_j}\(1+b_1x^\mu \partial_\mu +b_2x^\mu x^\nu \partial_\mu \partial_\nu+ \cdots \)\mathcal{O}_k(0)\, ,
\eea
where  $\Delta_i$, $\Delta_j$ and $\Delta_k$ are the scaling dimensions of their associated quasiprimary operators and the coefficients $C_{ijk}$ are dynamical parameters of the theory. We wrote the above expansion for scalar operators $\mathcal{O}_i$, $\mathcal{O}_j$ and $\mathcal{O}_k$, there are similar expansions for operators with arbitrary spin. In (\ref{OPE-expansion}), the coefficients $b_n$ depend only on the $\Delta_i$, $\Delta_j$, $\Delta_k$ and thus are determined by the kinematics of the conformal symmetry. The contribution that comes from a particular primary and its descendents is known as the OPE block, and we denote it by
\bea\label{OPE-block}
{\mathcal{B}}^{ij}_k(x,0)=|x|^{\Delta_k}\(1+b_1x^\mu \partial_\mu +b_2x^\mu x^\nu \partial_\mu \partial_\nu +\cdots \)\mathcal{O}_k(0)\,.
\eea
In particular, ${\mathcal{B}}^{ij}_k(x,0)$ becomes independent of $\Delta_i$ and $\Delta_j$ when $\mathcal{O}_i$, $\mathcal{O}_j$ have equal scaling dimension, $\Delta_i=\Delta_j$, thus in that case we denote it ${\mathcal{B}}_k(x,0)$. Hereafter, we will restrict to this situation for simplicity.

In Lorentzian signature, when $x_1,x_2$ are time-like separated, one can use the following integral representation for the OPE block 
\bea\label{OPE-int-rep}
{\mathcal{B}}_k(x_1,x_2)=c_k \int_{D(x_1,x_2)}\diff^d \xi \(\frac{|x_1-\xi ||x_2-\xi |}{|x_1-x_2|}\)^{\Delta_k-d}\mathcal{O}_k(\xi )\,.
\eea 
This expression can be derived from the shadow operator formalism \cite{Czech:2016xec}, \cite{SimmonsDuffin:2012uy}. In (\ref{OPE-int-rep}), $c_k$ is an arbitrary constant to be determined, and the space-time integral is taken over the causal diamond whose future and past causal tips are $x_1$ and $x_2$ respectively and denote it as $D(x_1,x_2)$. This formula can be written in a more compact form, in terms of the conformal Killing vector $K^\mu$ that preserves the causal diamond as
\bea\label{OPE-int-rep-2}
{\mathcal{B}}_k(x_1,x_2)=\frac{c_k}{\(2\pi\)^{\Delta_k-d}} \int_{D(x_1,x_2)}\diff^d \xi \, \, |K|^{\Delta_k-d} \mathcal{O}_k(\xi )\,,
\eea 
where 
\bea
K^\mu\partial_\mu=-\frac{2\pi }{(x_1-x_2)^2}\left[  (x_2-\xi)^2(x_1^\mu -\xi^\mu)  -(x_1-\xi)^2(x^\mu_2-\xi^\mu)\right]\partial_\mu\,, 
\eea
and 
\bea
|K|=2\pi \frac{|x_1-\xi||x_2-\xi|}{|x_1-x_2|}\,.
\eea
Indeed, (\ref{OPE-int-rep-2}) has a natural generalization for the OPE block of a spin $J$ symmetric quasiprimary operator $\mathcal{O}_k^{\mu_1 \mu_2 \cdots \mu_J}$  \cite{deBoer:2016pqk} which reads
\bea\label{OPE-int-rep-3}
{\mathcal{B}}_{k,J}(x_1,x_2)=\frac{c_k}{\(2\pi\)^{\Delta_k-d}} \int_{D(x_1,x_2)}\diff^d \xi \, \, |K|^{\Delta_k-d-J}K^{\mu_1}\cdots K^{\mu_J} \mathcal{O}_{k,\, \mu_1 \cdots \mu_J}(\xi )\,.
\eea 
The above expression can be further simplified for cases in which the operator $\mathcal{O}_{k,\, \mu_1 \cdots \mu_J}(\xi )$ corresponds to a traceless conserved spin $J$ operator. In that case $\Delta_k=J+d-2$ and we can construct the conserved current $J_\mu$ 
\bea\label{current-J}
J_\mu \equiv K^{\mu_2}\cdots K^{\mu_{J}}\mathcal{O}_{k,\, \mu \mu_2 \cdots \mu_J}(\xi )\,,
\eea 
where conservation of $J_{\mu}$ follows from the fact that $K^{\mu}$ is a conformal Killing vector, and that $\mathcal{O}_{k,\, \mu\mu_2 \cdots \mu_J}(\xi )$ is conserved and traceless. This results in the expression 
\bea\label{OPE-int-rep-4}
{\mathcal{B}}_{k,J}(x_1,x_2)=\frac{c_k}{\(2\pi\)^{J-2}} \int_{D(x_1,x_2)}\diff^d \xi \, \, \frac{K^\mu J_\mu }{|K|^{2}}\,.
\eea 
The space-time integral over the causal cone can be foliated with surfaces of constant $|K|$ so that $\diff^d\xi=\diff^{d-1}\Sigma |K|\diff \lambda $, where $\diff^{d-1}\Sigma$ is the area element induced on the constant $|K|$ surface and $\lambda$ is a flow parameter in the direction of $K^\mu$. In these coordinates, one realizes that the integral $\int \diff^{d-1}\Sigma \cdots$ is independent of $\lambda$ (by the conservation equation) and thus, one gets an overall divergent factor $\int \diff \lambda$ which can be absorbed into the coefficient $c_k$ by $c_k\to \tilde{c}_k$. Using the conservation of $J^\mu$ once more, one can show that the resulting integral over $\Sigma$ is indeed independent of the surface ---it is not any more subject to be a constant $|K|$ surface; see \cite{deBoer:2016pqk} for a more detailed explanation on this. This results in 
\bea\label{OPE-int-rep-5}
{\mathcal{B}}_{k,J}(x_1,x_2)=\frac{\tilde{c}_k}{\(2\pi\)^{J-2}} \int_{\Sigma_A}\diff^{d-1}\Sigma  \, \, n_A^\mu J_\mu \, ,
\eea
where $\Sigma_A$ is an arbitrary space-like surface with boundary on the sphere at the rim of the causal cone $\partial \Sigma_A=\mathbb S_A^{d-1}$.
On the other hand, the conformal block associated to a spin $J$ symmetric operator can be rewritten as the following correlator 
\bea
G^d_{\Delta_k,J}(u,v)=\langle{\mathcal{B}}_{k,J}(x_1,x_2){\mathcal{B}}_{k,J}(x_3,x_4) \rangle \,,
\eea
in terms of the associated OPE blocks.
This formula comes from the usual partial wave decomposition of a four point correlator and formulas (\ref{OPE-expansion}) and (\ref{OPE-block}). Thus, when such operator is conserved and traceless, one can use equation (\ref{OPE-int-rep-5}) in the above formula and get
\bea\label{G-emi}
G^d_{d-2+J,J}(u,v)=\frac{\tilde{c}^2_k}{\(2\pi\)^{2(J-2)}} \int_{\Sigma_A}\diff^{d-1}\Sigma_A  \, \, n_A^\mu \int_{\Sigma_B}\diff^{d-1}\Sigma_B  \, \, n_B^\nu  \langle J_\mu (\xi) J_\nu(\xi') \rangle \,,
\eea
where $J_\mu$ is given by (\ref{current-J}).
Furthermore, in the particular case of $J=1$,  $J_{\mu}$ becomes a quasiprimary conserved current operator since $J_\mu=\mathcal{O}_{k,\, \mu}$, and its two-point function is fixed by conformal invariance to be 
\bea\label{JJ}
\langle J_\mu (\xi) J_\nu (\xi')\rangle &=&\frac{1}{|\xi-\xi'|^{2(d-1)}}\(\eta_{\mu \nu}-\frac{2(\xi-\xi')_\mu (\xi-\xi')_\nu}{|\xi-\xi'|^2}  \)\,.
\eea
The above expression is obtained from the vector-vector correlator with 
$\Delta_k=d-1$. Further, (\ref{JJ}) can be rewritten as 
\bea\label{JJ-emi}
\langle J_\mu (\xi) J_\nu (\xi')\rangle &=&-\frac{1}{2(d-1)(d-2)}\(\partial_\mu \partial_\nu-g_{\mu \nu}\partial^2 \)\frac{1}{|\xi-\xi'|^{2(d-2)}}\,,
\eea
where $g_{\mu \nu}$ is the Minkowski metric. 
Finally, plugging (\ref{JJ-emi}) into (\ref{G-emi}) results in a formula for the conformal block of a spin $1$ conserved current that is identically proportional to the EMI formula as given in \req{aa}. 

\section{Conformal blocks for conserved currents \label{CB-CC}}

The authors of \cite{Hogervorst:2013sma} presented a general polynomial formula for the conformal blocks in any dimension as a function of its spin $J$ and conformal dimension $\Delta$.\footnote{We thank Dalimil Maz\' a\v{c} for point out this reference to us.} The expression has the form
\bea\label{Conf-Blocks-ap}
G^d_{\Delta,J}(u,v)=\sum_{n=0}^{\infty}|z|^{\Delta+n}\sum_{j}A_{n,j}\frac{C_j^{\frac{d-2}2}(\cos\varphi)}{C_j^{\frac{d-2}2}(1)}\, ,\quad A_{n,j}\geq 0\,,
\eea
where we expressed the $u$ and $v$ cross ratios in terms of the complex coordinates $z=|z|e^{i\varphi}$ with $\bar{z}=z^*$ via
\bea\label{u-zzbar}
u\equiv z\bar{z}, \qquad v\equiv (1-z)(1-\bar{z})\,.
\eea
$C_j^{\frac{d-2}2}(\cos\theta)$ are the Gegenbauer polynomials and the $j$ sum is over all the descendants of the primary operator $\Delta$ and thus it takes values
\bea
j=J+n, J+n-2, \dots, {\rm max}\(J-n, J+n \, {\rm mod} \, 2\)\, ,
\eea
for each $n$.
The coefficients $A_{n,j}$ are some universal functions of $\Delta, J$ and $d$ that are fixed by conformal symmetry. These coefficients obey some recursion relations that can be solved explicitly \cite{Hogervorst:2013sma}. In this section, we are interested in obtaining closed-form expressions for the simpler case in which the operator in question is a conserved current, and therefore its conformal dimension obeys
\bea
\Delta=J+d-2\, , \qquad J=0,1,2,\dots
\eea
In this case, it was found in \cite{Hogervorst:2013sma}, that only the maximal allowed spin $j=J+n$ has a nonzero coefficient. Such coefficient is given by 
\bea
A_{n,J+n}=\frac{(J+\frac {d-2}2)_n(J+d-2)_n}{n! (2J+d-2)_n}\,,
\eea
where $(x)_n=\Gamma(x+n)/\Gamma(x)$ are the Pochhammer coefficients. This fact simplifies considerably the formula (\ref{Conf-Blocks-ap}), which reduces to 
\bea\label{CB-app}
G^d_{d-2+J,J}(u,v)=\sum_{n=0}^{\infty}|z|^{d-2+J+n}\frac{(J+\frac {d-2}2)_n(J+d-2)_n}{n! (2J+d-2)_n}\frac{C_{J+n}^{\frac{d-2}2}(\cos\varphi)}{C_{J+n}^{\frac{d-2}2}(1)}\,.
\eea
The angular variable is given by 
\bea
\cos\varphi=\frac{z+\bar{z}}{2|z|}\, .
\eea
In particular, it is easy to check that for $\varphi=0$, which corresponds to the diagonal limit of the conformal blocks  $(z=\bar{z})$ the above expression reduces to
\bea
G^d_{d-2+J,J}(z)=z^{d-2+J}\,_2F_1\[J+\frac{d-2}2,J+d-2,2J+d-2,z\]\, ,
\eea
where we identified the Hypergeometric function in the series expansion of (\ref{CB-app}). This expression is indeed equivalent to the general expression for the diagonal limit of conformal blocks found in \cite{Hogervorst:2013kva} with $\Delta=J+d-2$. 

We would like to write down a similar integral representation to the one we obtained for the $J=1$ conformal block. In this case, it would be convenient to use the following integral representation of the Gegenbauer polynomials
\bea
C_{J+n}^{\frac{d-2}2}(\cos\varphi)=\frac{\Gamma\(J+n+d-2\)}{2^{d-3} n! \(\Gamma\!\({\frac{d-2}2}\)\)^2}\int_0^\pi  \diff\theta \, \sin^{d-3}\theta\(\cos\varphi+i \sin\varphi\cos \theta\)^{J+n}\,.
\eea  
The normalization coefficient is simply
\bea
C_{J+n}^{\frac{d-2}2}(1)=\frac{\Gamma\(J+n+d-2\)}{2^{d-3} n! \(\Gamma\!\({\frac{d-2}2}\)\)^2}\frac{\sqrt{\pi}\,\Gamma\!\(\frac{d-2}2\)}{\Gamma\!\(\frac{d-1}2\)}\,.
\eea
Plugging this into (\ref{CB-app}) and interchanging the integral with the convergent sum, one gets
\bea
G^d_{d-2+J,J}(u,v)&=&\frac{ \Gamma\!\(\frac{d-1}{2}\)}{\sqrt{\pi}\, \Gamma\!\(\frac{d-2}{2}\)}|z|^{d-2}\int_0^\pi  \diff \theta \, \sin^{d-3}\theta\, |z|^J\(\cos\varphi+i \sin\varphi \cos \theta\)^{J} \nonumber \\
&& \qquad \times \sum_{n=0}^{\infty}|z|^{n}\(\cos\varphi+i \sin\varphi \cos \theta\)^{n}\frac{(J+\frac {d-2}2)_n(J+d-2)_n}{n! (2J+d-2)_n}\, ,
\eea
where once again we recognize the sum over $n$ in the second line of the above equation as the series representation of a hypergeometric function, which leads to 
\bea\label{CB-int-3}
G^d_{d-2+J,J}(u,v)&=&\frac{2^{d-2} \, \Gamma\!\(\frac{d-1}2\)}{\sqrt{\pi}\,\Gamma\!\(\frac{d-2}{2}\)}|z|^{d-2}\int_0^{\pi/2}  \diff s \, \sin^{d-3}s\,\cos^{d-3}s \, \(z \cos^2s+\bar{z}\sin^2s \)^{J} \nonumber \\
&& \qquad \qquad \qquad \times\,_2F_1\[J+\frac{d-2}2,J+d-2,2J+d-2; z \cos^2s+\bar{z}\sin^2 s\]\,. \nonumber
\eea
We have additionally re-expressed the integral in terms of $z$ and $\bar{z}$ and changed the integration variable $t\to s/2$. Finally, we can get an integral with a bigger resemblance with (\ref{EMI-integral-3}) by doing $\cos s\to \xi$. In this way, we arrive at our final formula:
\bea\label{CB-int-4} 
G^d_{d-2+J,J}(u,v)&=&\frac{2^{d-2} \, \Gamma\!\(\frac{d-1}2\)}{\sqrt{\pi}\,\Gamma\!\(\frac{d-2}{2}\)}|z|^{d-2}\int_0^{1}  \diff \xi \,\(1-\xi^2\)^{\frac{d-4}{2}} \xi^{d-3} \, \(z \,\xi^2+\bar{z}\,(1-\xi^2) \)^{J}  \\
&& \qquad \qquad \qquad \times\,_2F_1\[J+\frac{d-2}2,J+d-2,2J+d-2; z \,\xi^2+\bar{z}\,(1-\xi^2) \]\,. \notag
\eea
To express it in terms of $u$ and $v$, one would need to invert the relations (\ref{u-zzbar}) and plug them into the above formula. However, in these variables, the explicit formula is not particularly illuminating.

Finally, for $J=1$ the above expression becomes   
\bea\label{current-CB}
G^d_{d-2+J,J}(u,v)&=&\frac{2^{d-2} \, \Gamma\!\(\frac{d-1}2\)}{\sqrt{\pi}\,\Gamma\!\(\frac{d-2}{2}\)}|z|^{d-2}\int_0^{1}  \diff \xi \,\(1-\xi^2\)^{\frac{d-4}{2}} \xi^{d-3} \, \(z \,\xi^2+\bar{z}\,(1-\xi^2) \) \nonumber \\
&& \qquad \qquad \qquad \times\,_2F_1\[\frac{d}2,d-1,d; z \,\xi^2+\bar{z}\,(1-\xi^2) \]\,. 
\eea
Unfortunately, it is not easy to analytically compare the above formula with (\ref{EMI-integral-3}).
Nevertheless, we checked numerically that indeed (\ref{current-CB}) is equivalent to (\ref{EMI-integral-3}), when written in the same variables. This completes our checks of the formulas presented in section \ref{313} particularly for $d=3$ and $d=5$.

\section{Coefficient of the first subleading term in the MI for a free fermion}
\label{coeff}
Here we want to compute the contribution of the two-copy primary $V_{\mu\nu}$ of eq. (\ref{ffff}) to the mutual information of a free fermion. 
Using the signature $(-1,1,\cdots,1)$ and normalizing the fermion field such that
\be
\langle \psi(0)\bar{\psi}(x)\rangle=i\frac{\slash \hspace{-.2cm}x}{|x|^{d}}\,,
\ee
the proportionality constant in the correlator (\ref{correla}) is $-4 \,d\, \textrm{tr}(1)$. 

The contribution to the mutual information is of the form 
\be
\Delta I(A,B)=\gamma\, n_A^\mu n_A^\nu\, \langle V_{\mu\nu}(0) V_{\alpha\beta}(L \,\hat{l})\rangle n_B^{\alpha} n_B^\beta=-\frac{4 d\, \textrm{tr}(1)\, \gamma}{L^{2 d}} \left((2(n_A\cdot l)(n_B\cdot l)-(n_A\cdot n_B))^2-\frac{1}{d}\right)\,,\label{siste}
\ee
where we seek to obtain the coefficient $\gamma$.

Following \cite{casini2021mutual}, this coefficient can be computed using the modular flow of spheres. We have
\bea
&& \gamma\, n_A^\alpha \,n_A^{\alpha\, '}\,n_B^\beta\, n_B^{\beta\,'}  \langle V_{\alpha \alpha '}(r_A) V_{\delta \delta '}(r_1)\rangle \langle V_{\beta \beta '}(r_B) V_{\eta \eta '}(r_2)\rangle  \label{319}\\
&&  =\int_{-\infty}^\infty \diff s\,\frac{\pi}{4 \cosh^2(\pi s)}\langle (\partial_\delta\bar{\psi}^{(s,A)}) \gamma_{\delta '} \psi-\bar{\psi}^{(s,A)} \gamma_{\delta '} \partial_\delta \psi+ (\partial_{\delta }\bar{\psi}) \gamma_{\delta '} \psi^{(s,A)}- \bar{\psi} \gamma_{\delta '} \partial_{\delta}\psi^{(s,A)}  \rangle (r_1)\,\nonumber\\ &&
\langle (\partial_\eta\bar{\psi}^{(s,B)}) \gamma_{\eta '} \psi-\bar{\psi}^{(s,B)} \gamma_{\eta '} \partial_\eta \psi+ (\partial_\eta\bar{\psi}) \gamma_{\eta '} \psi^{(s,B)}- \bar{\psi} \gamma_{\eta '} \partial_\eta\psi^{(s,B)}  \rangle (r_2)\,,\hspace{1cm} r_{1,2}^2\rightarrow \infty\,.\nonumber
\eea
 This formula gives the coefficient for any shape of $A,B$ in terms of the modular evolved fields $\psi^{(s,A)},\psi^{(s,B)}$. For a sphere with $n$ in the time direction the modular flow is given by the conformal transformation (see \cite{Haag:1992hx}) 
\bea
y^0(x,\tau) &=&  N(\tau)^{-1} R \left(x^0 R \cosh(2 \pi \tau)+\frac{1}{2} (R^2-x^2) \sinh(2 \pi \tau)\right) \nonumber   \,,\\
y^i(x,\tau) &=& N(\tau)^{-1} \, R^2\, x^i     \,,\\
N(\tau) &=&  x^0 R \sinh(2 \pi \tau)+\frac{1}{2} \cosh(2 \pi \tau) (R^2- x^2)+\frac{1}{2}(R^2+ x^2) \nonumber\,,
\eea
with $x^2=-(x^0)^2+x^i x^i$. 
The modular evolved field is
\be
\psi^{(\tau,A)}(x)= \Omega^{(d-1)/2}(x,\tau) S(\Lambda(x,\tau))^{-1} \psi(y(x,\tau))\,,
\ee
where the Lorentz transformation $\Lambda$ and conformal factor $\Omega$ follow from
\be
\Lambda^\mu_\nu= \Omega^{-1} \, \frac{\diff y^\mu}{\diff  x^\nu}\,. 
\ee
$\Lambda$ is a Lorentz boost in the direction of $\vec{x}$ of a certain hyperbolic angle $\nu(\tau,x)$ given by
\be
\sinh(\nu(\tau,x))=-\frac{4 \sinh(\pi \tau) |\vec{x}| (R \cosh(\pi \tau)+x^0 \sinh(\pi\tau) )}{R^2+2 R x^0 \sinh(2 \pi\tau) + x^2 +(R^2-x^2) \cosh(2 \pi\tau)}\,.
\ee
 The representation $S(\Lambda)$ can be obtained by the formula 
\be
S=\cosh(\nu/2) \, 1- \gamma^0 \gamma^i \hat{x}_i \, \sinh(\nu/2)\,.\label{maty}
\ee

In formula (\ref{319}) we have to choose a complex modular parameter for the transformation as
\be
\tau= s+ i/2\,,
\ee
and take $r_1\cdot n_A=r_2\cdot n_B=0$. We can use local coordinate systems for each sphere where $n_A, n_B$ are in the time direction for each of the correlators.    

The right hand side of (\ref{319}) gives
\be
\textrm{tr}(1)^2 \frac{2 \sqrt{\pi} d \, \Gamma\!\(d+1\)}{\Gamma\!\(d+\frac{3}{2}\)} \left(n_A^\delta n_A^{\delta '}+ \frac{1}{d} \,g^{\delta \delta '}\right)\,\left(n_B^\eta n_B^{\eta '}+ \frac{1}{d} \,g^{\eta \eta '}\right)\ \frac{R_A^d R_B^d}{r_1^{2 d} r_2^{2 d} }\,.
\ee
 From which we obtain the coefficient
\be
\gamma=  \frac{ \sqrt{\pi}\, \Gamma\!\(d+1\)}{8 d\, \Gamma\!\(d+\frac32\)}\, R_A^d R_B^d\,.
\ee
Upon replacing in (\ref{siste}), the coefficient of the contribution (\ref{hh}) in the main text follows.

\bibliography{EE}{}

\providecommand{\href}[2]{#2}\begingroup\raggedright\begin{thebibliography}{10}

\bibitem{streater2000pct}
R.~F. Streater and A.~S. Wightman, {\em PCT, spin and statistics, and all
  that}, vol.~52.
\newblock Princeton University Press, 2000.

\bibitem{Haag:1992hx}
R.~Haag, {\em {Local quantum physics: Fields, particles, algebras}}.
\newblock
1992.
\newblock
%%CITATION = INSPIRE-338216;%%.

\bibitem{Cardy.esferaslejanas}
J.~Cardy, ``Some results on the mutual information of disjoint regions in
  higher dimensions,'' {\em Journal of Physics A: Mathematical and Theoretical}
  {\bfseries 46} no.~28, (2013) 285402.
  \url{http://stacks.iop.org/1751-8121/46/i=28/a=285402}.

\bibitem{agon2016quantum}
C.~A. Ag{\'o}n and T.~Faulkner, ``Quantum corrections to holographic mutual
  information,'' {\em Journal of High Energy Physics} {\bfseries 2016} no.~8,
  (2016) 118.

\bibitem{tobe}
C.~Ag\'on, P.~Bueno, and H.~Casini, ``{To appear},''.

\bibitem{ohya2004quantum}
M.~Ohya and D.~Petz, {\em Quantum entropy and its use}.
\newblock Springer Science \& Business Media, 2004.

\bibitem{casini2021mutual}
H.~Casini, E.~Test{\'e}, and G.~Torroba, ``Mutual information superadditivity
  and unitarity bounds,'' {\em arXiv preprint arXiv:2103.15847} (2021) .

\bibitem{casini2017modular}
H.~Casini, E.~Test{\'e}, and G.~Torroba, ``Modular hamiltonians on the null
  plane and the markov property of the vacuum state,'' {\em Journal of Physics
  A: Mathematical and Theoretical} {\bfseries 50} no.~36, (2017) 364001.

\bibitem{casini2018all}
H.~Casini, E.~Test{\'e}, and G.~Torroba, ``All the entropies on the
  light-cone,'' {\em Journal of High Energy Physics} {\bfseries 2018} no.~5,
  (2018) 5.

\bibitem{casini2010entropy}
H.~Casini, ``Entropy inequalities from reflection positivity,'' {\em Journal of
  Statistical Mechanics: Theory and Experiment} {\bfseries 2010} no.~08, (2010)
  P08019.

\bibitem{ryu2006holographic}
S.~Ryu and T.~Takayanagi, ``Holographic derivation of entanglement entropy from
  the anti--de sitter space/conformal field theory correspondence,'' {\em
  Physical review letters} {\bfseries 96} no.~18, (2006) 181602.

\bibitem{Ryu:2006ef}
S.~Ryu and T.~Takayanagi, ``{Aspects of Holographic Entanglement Entropy},''
  \href{http://dx.doi.org/10.1088/1126-6708/2006/08/045}{{\em JHEP} {\bfseries
  08} (2006) 045}, \href{http://arxiv.org/abs/hep-th/0605073}{{\ttfamily
  arXiv:hep-th/0605073}}.

\bibitem{Hubeny:2007xt}
V.~E. Hubeny, M.~Rangamani, and T.~Takayanagi, ``{A Covariant holographic
  entanglement entropy proposal},''
  \href{http://dx.doi.org/10.1088/1126-6708/2007/07/062}{{\em JHEP} {\bfseries
  07} (2007) 062}, \href{http://arxiv.org/abs/0705.0016}{{\ttfamily
  arXiv:0705.0016 [hep-th]}}.

\bibitem{Lewkowycz:2013nqa}
A.~Lewkowycz and J.~Maldacena, ``{Generalized gravitational entropy},''
  \href{http://dx.doi.org/10.1007/JHEP08(2013)090}{{\em JHEP} {\bfseries 08}
  (2013) 090}, \href{http://arxiv.org/abs/1304.4926}{{\ttfamily arXiv:1304.4926
  [hep-th]}}.

\bibitem{Faulkner:2013ana}
T.~Faulkner, A.~Lewkowycz, and J.~Maldacena, ``{Quantum corrections to
  holographic entanglement entropy},''
  \href{http://dx.doi.org/10.1007/JHEP11(2013)074}{{\em JHEP} {\bfseries 11}
  (2013) 074}, \href{http://arxiv.org/abs/1307.2892}{{\ttfamily arXiv:1307.2892
  [hep-th]}}.

\bibitem{Casini:2008wt}
H.~Casini and M.~Huerta, ``{Remarks on the entanglement entropy for
  disconnected regions},''
  \href{http://dx.doi.org/10.1088/1126-6708/2009/03/048}{{\em JHEP} {\bfseries
  03} (2009) 048}, \href{http://arxiv.org/abs/0812.1773}{{\ttfamily
  arXiv:0812.1773 [hep-th]}}.

\bibitem{casini2015mutual}
H.~Casini, M.~Huerta, R.~C. Myers, and A.~Yale, ``Mutual information and the
  f-theorem,'' {\em Journal of High Energy Physics} {\bfseries 2015} no.~10,
  (2015) 1--70.

\bibitem{Bueno:2015rda}
P.~Bueno, R.~C. Myers, and W.~Witczak-Krempa, ``{Universality of corner
  entanglement in conformal field theories},''
  \href{http://dx.doi.org/10.1103/PhysRevLett.115.021602}{{\em Phys. Rev.
  Lett.} {\bfseries 115} (2015) 021602},
  \href{http://arxiv.org/abs/1505.04804}{{\ttfamily arXiv:1505.04804
  [hep-th]}}.

\bibitem{Witczak-Krempa:2016jhc}
W.~Witczak-Krempa, L.~E. Hayward~Sierens, and R.~G. Melko, ``{Cornering gapless
  quantum states via their torus entanglement},''
  \href{http://dx.doi.org/10.1103/PhysRevLett.118.077202}{{\em Phys. Rev.
  Lett.} {\bfseries 118} no.~7, (2017) 077202},
  \href{http://arxiv.org/abs/1603.02684}{{\ttfamily arXiv:1603.02684
  [cond-mat.str-el]}}.

\bibitem{Bueno:2019mex}
P.~Bueno, H.~Casini, and W.~Witczak-Krempa, ``{Generalizing the entanglement
  entropy of singular regions in conformal field theories},''
  \href{http://dx.doi.org/10.1007/JHEP08(2019)069}{{\em JHEP} {\bfseries 08}
  (2019) 069}, \href{http://arxiv.org/abs/1904.11495}{{\ttfamily
  arXiv:1904.11495 [hep-th]}}.

\bibitem{Estienne:2021lxh}
B.~Estienne, J.-M. St\'ephan, and W.~Witczak-Krempa, ``{Cornering the universal
  shape of fluctuations},'' \href{http://arxiv.org/abs/2102.06223}{{\ttfamily
  arXiv:2102.06223 [cond-mat.str-el]}}.

\bibitem{casini2005entanglement}
H.~Casini, C.~Fosco, and M.~Huerta, ``Entanglement and alpha entropies for a
  massive dirac field in two dimensions,'' {\em Journal of Statistical
  Mechanics: Theory and Experiment} {\bfseries 2005} no.~07, (2005) P07007.

\bibitem{swingle2010mutual}
B.~Swingle, ``Mutual information and the structure of entanglement in quantum
  field theory,'' {\em arXiv preprint arXiv:1010.4038} (2010) .

\bibitem{Hislop:1981uh}
P.~D. Hislop and R.~Longo, ``{Modular Structure of the Local Algebras
  Associated With the Free Massless Scalar Field Theory},''
\href{http://dx.doi.org/10.1007/BF01208372}{{\em Commun. Math. Phys.}
  {\bfseries 84} (1982) 71}.
%%CITATION = CMPHA,84,71;%%.

\bibitem{chen2017mutual}
B.~Chen, L.~Chen, P.-x. Hao, and J.~Long, ``On the mutual information in
  conformal field theory,'' {\em Journal of High Energy Physics} {\bfseries
  2017} no.~6, (2017) 96.

\bibitem{Hofman:2008ar}
D.~M. Hofman and J.~Maldacena, ``{Conformal collider physics: Energy and charge
  correlations},'' \href{http://dx.doi.org/10.1088/1126-6708/2008/05/012}{{\em
  JHEP} {\bfseries 05} (2008) 012},
\href{http://arxiv.org/abs/0803.1467}{{\ttfamily arXiv:0803.1467 [hep-th]}}.
%%CITATION = ARXIV:0803.1467;%%.

\bibitem{Hofman:2016awc}
D.~M. Hofman, D.~Li, D.~Meltzer, D.~Poland, and F.~Rejon-Barrera, ``{A Proof of
  the Conformal Collider Bounds},''
  \href{http://dx.doi.org/10.1007/JHEP06(2016)111}{{\em JHEP} {\bfseries 06}
  (2016) 111}, \href{http://arxiv.org/abs/1603.03771}{{\ttfamily
  arXiv:1603.03771 [hep-th]}}.

\bibitem{casini2009reduced}
H.~Casini and M.~Huerta, ``Reduced density matrix and internal dynamics for
  multicomponent regions,'' {\em Classical and quantum gravity} {\bfseries 26}
  no.~18, (2009) 185005.

\bibitem{Casini:2009sr}
H.~Casini and M.~Huerta, ``{Entanglement entropy in free quantum field
  theory},'' \href{http://dx.doi.org/10.1088/1751-8113/42/50/504007}{{\em J.
  Phys. A} {\bfseries 42} (2009) 504007},
  \href{http://arxiv.org/abs/0905.2562}{{\ttfamily arXiv:0905.2562 [hep-th]}}.

\bibitem{Klebanov:2011gs}
I.~R. Klebanov, S.~S. Pufu, and B.~R. Safdi, ``{F-Theorem without
  Supersymmetry},'' \href{http://dx.doi.org/10.1007/JHEP10(2011)038}{{\em JHEP}
  {\bfseries 10} (2011) 038}, \href{http://arxiv.org/abs/1105.4598}{{\ttfamily
  arXiv:1105.4598 [hep-th]}}.

\bibitem{Marino:2011nm}
M.~Marino, ``{Lectures on localization and matrix models in supersymmetric
  Chern-Simons-matter theories},''
  \href{http://dx.doi.org/10.1088/1751-8113/44/46/463001}{{\em J. Phys. A}
  {\bfseries 44} (2011) 463001},
  \href{http://arxiv.org/abs/1104.0783}{{\ttfamily arXiv:1104.0783 [hep-th]}}.

\bibitem{Casini:2006hu}
H.~Casini and M.~Huerta, ``{Universal terms for the entanglement entropy in 2+1
  dimensions},'' \href{http://dx.doi.org/10.1016/j.nuclphysb.2006.12.012}{{\em
  Nucl. Phys. B} {\bfseries 764} (2007) 183--201},
  \href{http://arxiv.org/abs/hep-th/0606256}{{\ttfamily arXiv:hep-th/0606256}}.

\bibitem{Hirata:2006jx}
T.~Hirata and T.~Takayanagi, ``{AdS/CFT and strong subadditivity of
  entanglement entropy},''
  \href{http://dx.doi.org/10.1088/1126-6708/2007/02/042}{{\em JHEP} {\bfseries
  02} (2007) 042}, \href{http://arxiv.org/abs/hep-th/0608213}{{\ttfamily
  arXiv:hep-th/0608213}}.

\bibitem{Swingle:2010jz}
B.~Swingle, ``{Mutual information and the structure of entanglement in quantum
  field theory},'' \href{http://arxiv.org/abs/1010.4038}{{\ttfamily
  arXiv:1010.4038 [quant-ph]}}.

\bibitem{Osborn:1993cr}
H.~Osborn and A.~C. Petkou, ``{Implications of conformal invariance in field
  theories for general dimensions},''
  \href{http://dx.doi.org/10.1006/aphy.1994.1045}{{\em Annals Phys.} {\bfseries
  231} (1994) 311--362}, \href{http://arxiv.org/abs/hep-th/9307010}{{\ttfamily
  arXiv:hep-th/9307010}}.

\bibitem{Faulkner:2015csl}
T.~Faulkner, R.~G. Leigh, and O.~Parrikar, ``{Shape Dependence of Entanglement
  Entropy in Conformal Field Theories},''
  \href{http://dx.doi.org/10.1007/JHEP04(2016)088}{{\em JHEP} {\bfseries 04}
  (2016) 088},
\href{http://arxiv.org/abs/1511.05179}{{\ttfamily arXiv:1511.05179 [hep-th]}}.
%%CITATION = ARXIV:1511.05179;%%.

\bibitem{Solodukhin:2008dh}
S.~N. Solodukhin, ``{Entanglement entropy, conformal invariance and extrinsic
  geometry},'' \href{http://dx.doi.org/10.1016/j.physletb.2008.05.071}{{\em
  Phys. Lett. B} {\bfseries 665} (2008) 305--309},
  \href{http://arxiv.org/abs/0802.3117}{{\ttfamily arXiv:0802.3117 [hep-th]}}.

\bibitem{Dowker:1976zf}
J.~S. Dowker and R.~Critchley, ``{The Stress Tensor Conformal Anomaly for
  Scalar and Spinor Fields},''
  \href{http://dx.doi.org/10.1103/PhysRevD.16.3390}{{\em Phys. Rev. D}
  {\bfseries 16} (1977) 3390}.

\bibitem{Duff:1977ay}
M.~J. Duff, ``{Observations on Conformal Anomalies},''
  \href{http://dx.doi.org/10.1016/0550-3213(77)90410-2}{{\em Nucl. Phys. B}
  {\bfseries 125} (1977) 334--348}.

\bibitem{Bastianelli:2000hi}
F.~Bastianelli, S.~Frolov, and A.~A. Tseytlin, ``{Conformal anomaly of (2,0)
  tensor multiplet in six-dimensions and AdS / CFT correspondence},''
  \href{http://dx.doi.org/10.1088/1126-6708/2000/02/013}{{\em JHEP} {\bfseries
  02} (2000) 013}, \href{http://arxiv.org/abs/hep-th/0001041}{{\ttfamily
  arXiv:hep-th/0001041}}.

\bibitem{Safdi:2012sn}
B.~R. Safdi, ``{Exact and Numerical Results on Entanglement Entropy in
  (5+1)-Dimensional CFT},''
  \href{http://dx.doi.org/10.1007/JHEP12(2012)005}{{\em JHEP} {\bfseries 12}
  (2012) 005}, \href{http://arxiv.org/abs/1206.5025}{{\ttfamily arXiv:1206.5025
  [hep-th]}}.

\bibitem{Long:2016vkg}
J.~Long, ``{On co-dimension two defect operators},''
  \href{http://arxiv.org/abs/1611.02485}{{\ttfamily arXiv:1611.02485
  [hep-th]}}.

\bibitem{Chen:2016mya}
B.~Chen and J.~Long, ``{R\'enyi mutual information for a free scalar field in
  even dimensions},'' \href{http://dx.doi.org/10.1103/PhysRevD.96.045006}{{\em
  Phys. Rev. D} {\bfseries 96} no.~4, (2017) 045006},
  \href{http://arxiv.org/abs/1612.00114}{{\ttfamily arXiv:1612.00114
  [hep-th]}}.

\bibitem{Chen:2017hbk}
B.~Chen, L.~Chen, P.-x. Hao, and J.~Long, ``{On the Mutual Information in
  Conformal Field Theory},''
  \href{http://dx.doi.org/10.1007/JHEP06(2017)096}{{\em JHEP} {\bfseries 06}
  (2017) 096}, \href{http://arxiv.org/abs/1704.03692}{{\ttfamily
  arXiv:1704.03692 [hep-th]}}.

\bibitem{Dolan:2011dv}
F.~A. Dolan and H.~Osborn, ``{Conformal Partial Waves: Further Mathematical
  Results},'' \href{http://arxiv.org/abs/1108.6194}{{\ttfamily arXiv:1108.6194
  [hep-th]}}.

\bibitem{Hogervorst:2013kva}
M.~Hogervorst, H.~Osborn, and S.~Rychkov, ``{Diagonal Limit for Conformal
  Blocks in $d$ Dimensions},''
  \href{http://dx.doi.org/10.1007/JHEP08(2013)014}{{\em JHEP} {\bfseries 08}
  (2013) 014}, \href{http://arxiv.org/abs/1305.1321}{{\ttfamily arXiv:1305.1321
  [hep-th]}}.

\bibitem{ElShowk:2012ht}
S.~El-Showk, M.~F. Paulos, D.~Poland, S.~Rychkov, D.~Simmons-Duffin, and
  A.~Vichi, ``{Solving the 3D Ising Model with the Conformal Bootstrap},''
  \href{http://dx.doi.org/10.1103/PhysRevD.86.025022}{{\em Phys. Rev. D}
  {\bfseries 86} (2012) 025022},
  \href{http://arxiv.org/abs/1203.6064}{{\ttfamily arXiv:1203.6064 [hep-th]}}.

\bibitem{Hogervorst:2013sma}
M.~Hogervorst and S.~Rychkov, ``{Radial Coordinates for Conformal Blocks},''
  \href{http://dx.doi.org/10.1103/PhysRevD.87.106004}{{\em Phys. Rev. D}
  {\bfseries 87} (2013) 106004},
  \href{http://arxiv.org/abs/1303.1111}{{\ttfamily arXiv:1303.1111 [hep-th]}}.

\bibitem{Dolan:2000ut}
F.~A. Dolan and H.~Osborn, ``{Conformal four point functions and the operator
  product expansion},''
  \href{http://dx.doi.org/10.1016/S0550-3213(01)00013-X}{{\em Nucl. Phys. B}
  {\bfseries 599} (2001) 459--496},
  \href{http://arxiv.org/abs/hep-th/0011040}{{\ttfamily arXiv:hep-th/0011040}}.

\bibitem{shiba2020direct}
N.~Shiba, ``Direct calculation of mutual information of distant regions,'' {\em
  Journal of High Energy Physics} {\bfseries 2020} no.~9, (2020) 1--22.

\bibitem{epstein1963borchers}
H.~Epstein, ``On the borchers class of a free field,'' {\em Nuovo Cimento}
  {\bfseries 27} no.~CERN-TH-296, (1963) 886--893.

\bibitem{Casini:2015dsg}
H.~Casini and M.~Huerta, ``{Entanglement entropy of a Maxwell field on the
  sphere},'' \href{http://dx.doi.org/10.1103/PhysRevD.93.105031}{{\em Phys.
  Rev. D} {\bfseries 93} no.~10, (2016) 105031},
  \href{http://arxiv.org/abs/1512.06182}{{\ttfamily arXiv:1512.06182
  [hep-th]}}.

\bibitem{arias2018entropy}
R.~E. Arias, H.~Casini, M.~Huerta, and D.~Pontello, ``Entropy and modular
  hamiltonian for a free chiral scalar in two intervals,'' {\em Physical Review
  D} {\bfseries 98} no.~12, (2018) 125008.

\bibitem{casini2012positivity}
H.~Casini and M.~Huerta, ``Positivity, entanglement entropy, and minimal
  surfaces,'' {\em Journal of High Energy Physics} {\bfseries 2012} no.~11,
  (2012) 87.

\bibitem{blanco2019Renyi}
D.~Blanco, L.~Lanosa, M.~Leston, and G.~P{\'e}rez-Nadal, ``R{\'e}nyi mutual
  information inequalities from rindler positivity,'' {\em Journal of High
  Energy Physics} {\bfseries 2019} no.~12, (2019) 1--17.

\bibitem{Casini:2006ws}
H.~Casini, ``{Mutual information challenges entropy bounds},''
  \href{http://dx.doi.org/10.1088/0264-9381/24/5/013}{{\em Class. Quant. Grav.}
  {\bfseries 24} (2007) 1293--1302},
  \href{http://arxiv.org/abs/gr-qc/0609126}{{\ttfamily arXiv:gr-qc/0609126}}.

\bibitem{Myers:2010xs}
R.~C. Myers and A.~Sinha, ``{Seeing a c-theorem with holography},''
  \href{http://dx.doi.org/10.1103/PhysRevD.82.046006}{{\em Phys. Rev.}
  {\bfseries D82} (2010) 046006},
\href{http://arxiv.org/abs/1006.1263}{{\ttfamily arXiv:1006.1263 [hep-th]}}.
%%CITATION = ARXIV:1006.1263;%%.

\bibitem{Duff:1993wm}
M.~J. Duff, ``{Twenty years of the Weyl anomaly},''
  \href{http://dx.doi.org/10.1088/0264-9381/11/6/004}{{\em Class. Quant. Grav.}
  {\bfseries 11} (1994) 1387--1404},
  \href{http://arxiv.org/abs/hep-th/9308075}{{\ttfamily arXiv:hep-th/9308075}}.

\bibitem{Casini:2011kv}
H.~Casini, M.~Huerta, and R.~C. Myers, ``{Towards a derivation of holographic
  entanglement entropy},''
  \href{http://dx.doi.org/10.1007/JHEP05(2011)036}{{\em JHEP} {\bfseries 05}
  (2011) 036},
\href{http://arxiv.org/abs/1102.0440}{{\ttfamily arXiv:1102.0440 [hep-th]}}.
%%CITATION = ARXIV:1102.0440;%%.

\bibitem{Czech:2016xec}
B.~Czech, L.~Lamprou, S.~McCandlish, B.~Mosk, and J.~Sully, ``{A Stereoscopic
  Look into the Bulk},'' \href{http://dx.doi.org/10.1007/JHEP07(2016)129}{{\em
  JHEP} {\bfseries 07} (2016) 129},
  \href{http://arxiv.org/abs/1604.03110}{{\ttfamily arXiv:1604.03110
  [hep-th]}}.

\bibitem{deBoer:2016pqk}
J.~de~Boer, F.~M. Haehl, M.~P. Heller, and R.~C. Myers, ``{Entanglement,
  holography and causal diamonds},''
  \href{http://dx.doi.org/10.1007/JHEP08(2016)162}{{\em JHEP} {\bfseries 08}
  (2016) 162}, \href{http://arxiv.org/abs/1606.03307}{{\ttfamily
  arXiv:1606.03307 [hep-th]}}.

\bibitem{SimmonsDuffin:2012uy}
D.~Simmons-Duffin, ``{Projectors, Shadows, and Conformal Blocks},''
  \href{http://dx.doi.org/10.1007/JHEP04(2014)146}{{\em JHEP} {\bfseries 04}
  (2014) 146}, \href{http://arxiv.org/abs/1204.3894}{{\ttfamily arXiv:1204.3894
  [hep-th]}}.

\end{thebibliography}\endgroup
\bibliographystyle{utphys}

\end{document}